\documentclass[journal=jacsat,manuscript=article]{achemso}

\usepackage[version=3]{mhchem} 
\usepackage{graphicx}
\usepackage{bm}
\usepackage{lscape}
\usepackage{amsmath}
\usepackage{amssymb}
\usepackage{braket}
\usepackage{longtable}
\usepackage{array}
\usepackage{natbib}
\usepackage{xcolor} 
\usepackage{epstopdf}
\SectionNumbersOn

\author{Jun Jiang}
\email{jiang12@llnl.gov}
\phone{925-424-4296}
\affiliation[Lawrence Livermore National Laboratory]
{Center for Accelerator Mass Spectrometry, Lawrence Livermore National Laboratory, Livermore, California 94550, USA}

\title
  {The diabatic valence-hole concept}

\begin{document}

\begin{tocentry}
\center
\includegraphics{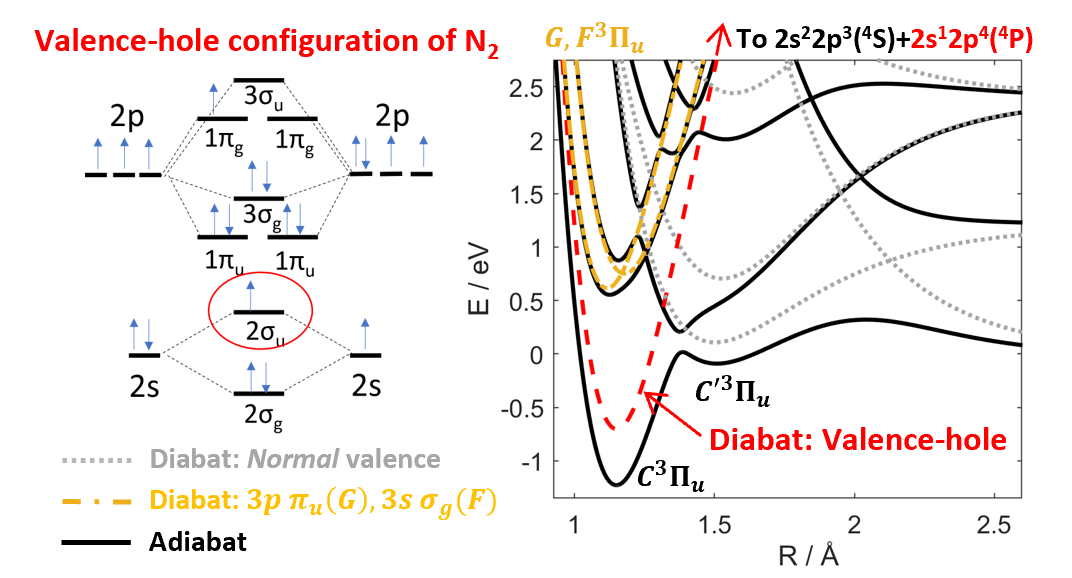}

\end{tocentry}
\newpage
\begin{abstract}

A global diabatization scheme, based on the ``valence-hole'' concept, has been previously applied to model webs of avoided-crosssings that exist in four electronic-state symmetry manifolds of C$_2$ ($^1\Pi_g$, $^3\Pi_g$, $^1\Sigma_u^+$, $^3\Sigma_u^+$). Here, this model is extended to the electronically excited states of four more molecules: CN ($^2\Sigma^+$), N$_2$ ($^3\Pi_u$), SiC ($^3\Pi$), and Si$_2$ ($^3\Pi_g$). Many strangenesses in the spectroscopic observations (e.g., energy level structure, predissociation linewidths, and radiative lifetimes) for all four electronic state systems discussed here are accounted for by this $unified$ model. The key concept of the model is valence-hole electron configurations: $3\sigma^24\sigma1\pi^45\sigma^2$ in CN ($^2\Sigma^+$), $2\sigma_g^22\sigma_u^11\pi_{u}^43\sigma_g^21\pi_{g}^1$ in N$_2$ ($^3\Pi_u$), $5\sigma^26\sigma7\sigma^22\pi^3$ in SiC ($^3\Pi$), and $4\sigma_g^24\sigma_u^15\sigma_g^22\pi_{u}^3$ in Si$_2$ ($^3\Pi_g$). These valence-hole configurations have a nominal bond order of three or higher, and correlate with high-energy separated-atom limits with an np$\leftarrow$ns (n=2,\,3) promotion in $one$ of the atomic constituents. This promotion results in a triply-occupied ``valence-core" (i.e., $2\sigma_g^22\sigma_u^1$ or the equivalent). On its way to dissociation, the strongly-bound diabatic valence-hole state crosses multiple weakly-bound or repulsive states, which belong to electron configurations with a completely-filled valence-core. These curve-crossings between diabatic potentials result in a network of many avoided-crossings among multiple electronic states, analogous to the well-studied electronic structure landscape of ionic-covalent crossings in strongly ionic molecules. Considering the unique role of valence-hole states in shaping the global electronic structure, the valence-hole concept should be added to our intuitive framework of chemical bonding.

\end{abstract}


\newpage
\section{Introduction}
\label{sec:intro}
Diatomic molecules are fundamental in shaping our intuitive understanding of electronic structure theory. Despite their structural simplicity, chemical bonding mechanisms (i.e., from the separated-atom limit to the equilibrium internuclear separation, $R_e$) are well understood only for a few lowest energy electronic states, even for the six C/N/O diatomic molecules. Pervasive configuration interactions and lumpy adiabatic potentials are prominent features of most electronically excited states. Detailed understanding and clear physical interpretation of chemical bond formation becomes increasingly challenging for higher energy electronic states, because of the reduced utility of simple molecular orbital (MO) concepts such as bond order and dissociation correlation diagram. 

A $global$ diabatization scheme is applied here.  This scheme is based on the existence of strongly-bound ``valence-hole" states. It is the foundation for a model of the electronic structure of selected electronic-state symmetry manifolds of several second- and third-row diatomic species: N$_2$ $^3\Pi_u$, CN $^2\Sigma^+$, Si$_2$ $^3\Pi_g$, and SiC $^3\Pi$. This diabatic representation of the electronic structure was originally applied to four electronic symmetry species of the C$_2$ molecule ($^1\Pi_g$, $^3\Pi_g$, $^1\Sigma_u^+$, and $^3\Sigma_u^+$)~\cite{{jiang2022diabatic},{Borsovszky2021}}. As my co-workers and I have revealed in C$_2$, the presence of $one$ valence-hole state fundamentally alters the global electronic structure landscape of each of the four systems of electronically excited states investigated here. The hitherto largely neglected valence-hole concept provides a $unified$ framework for understanding and exploiting diverse strangenesses of every electronic state in these electronic state systems: unusually-shaped adiabatic potentials, ``rule''-breaking energy level structure, and strongly $R$-dependent spectroscopic proprieties. I propose here the addition of the valence-hole concept to the intuitive framework of electronic structure.



\begin{figure}
\includegraphics[width=6.4 in]{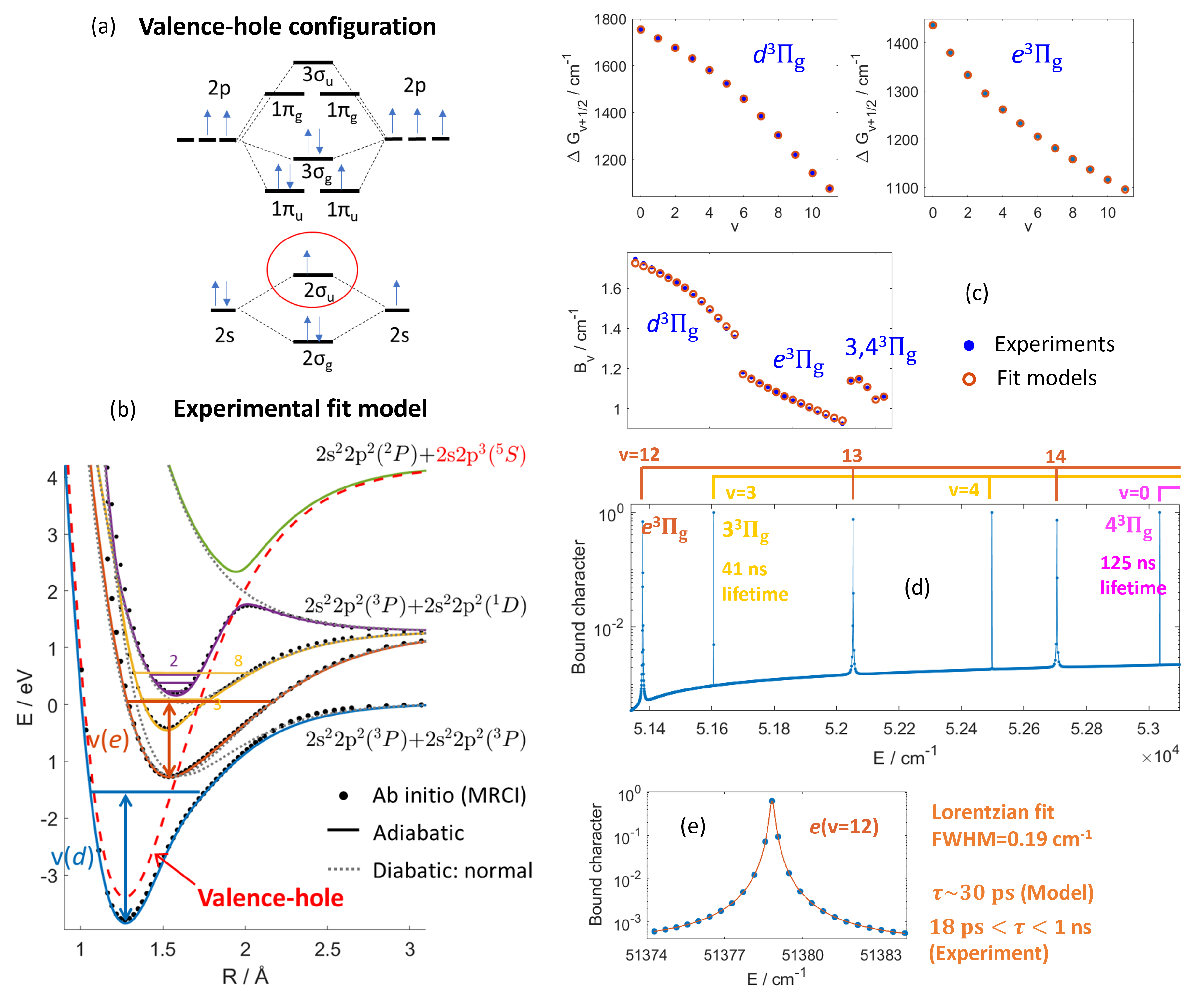}
\caption{Valence-hole model for the $^3\Pi_g$ electronic states of C$_2$, based on the fit to the experimental observations. (a) Valence-hole configuration at the $R_e$ of the $d^3\Pi_g$ state. (b) Diabatic and adiabatic potentials from the experimental fit model. The $ab$ $initio$ potential energies calculated with the multi-reference configuration interaction (MRCI) method are shown for comparison. (c) Comparison of the observed vibration-rotation constants of the C$_2$ $^3\Pi_g$ electronic states with the fit model output. The vibronic term energies and the rotational $B$ constants are reproduced, respectively, with a mean absolute error of 0.34 cm$^{-1}$  and 0.0077 cm$^{-1}$. (d) Bound diabatic basis state character ($J=1$) as a function of energy above the first dissociation limit of C$_2$. As a result of bound-continuum interaction, the bound character spreads into the nearby continuum. For each resonance, the predissociation lifetime is inversely proportional to the full-width half-maximum (FWHM) of the spread. The observed predissociation lifetimes for $3^3\Pi_g$ $v=3$ and $4^3\Pi_g$ $v=0$ from Ref~\citenum{krechkivska2017first} are indicated on the figure. (e) Bound diabatic basis state character near the $e^3\Pi_g$ $v=12$ resonance. The predissociation lifetime for the $e$-state $v=12$ level was determined in Ref~\citenum{welsh2017pi}.}
\label{fig:c2_3pig}
\end{figure}

The lowest-energy electronic state in each of the four electronic symmetry manifolds of C$_2$ ($^1\Pi_g$, $^3\Pi_g$, $^1\Sigma_u^+$, and $^3\Sigma_u^+$) is dominated by a valence-hole electron configuration at $R_e$. In these valence-hole configurations, $2\sigma_g^22\sigma_u^11\pi_{u}^33\sigma_g^2$ for $^{1,3}\Pi_g$ states and $2\sigma_g^22\sigma_u^11\pi_{u}^43\sigma_g^1$ for $^{1,3}\Sigma_u^+$ states, an electron has been promoted from the nominally anti-bonding $2\sigma_u$ MO to one of the bonding MOs of C$_2$ (see Fig.~\ref{fig:c2_3pig}a for the $^3\Pi_g$ valence-hole configuration of C$_2$). These valence-hole configurations have a nominal bond order of three, and correlate with a 2s$^2$2p$^2$+2s2p$^3$ separated-atom limit. The separated-atom correlation is a consequence of the triply-occupied ``valence-core" (i.e., $2\sigma_g^22\sigma_u^1$) in a valence-hole configuration. On its way to dissociation, the strongly-bound diabatic valence-hole state crosses multiple weakly-bound or repulsive states that are composed of electron configurations with a ``normal'' $2\sigma_g^22\sigma_u^2$ valence-core. These curve-crossings lead to a series of avoided-crossings among consecutive same-symmetry electronic states, analogous to the well-studied electronic structure of ionic (A$^+$B$^-$)-covalent (AB) crossings in strongly ionic molecules~\cite{herzberg1950molecular,london1932theorie,zewail2000femtochemistry}. In both cases, the global electronic structure landscape is fundamentally altered by a $single$ strongly-bound diabatic electronic state (ionic or valence-hole), which dissociates into a distinctly different, higher-energy channel (A$^+$+B$^-$ or 2s$^2$2p$^2$+2s2p$^3$) than all the other states in the same energy region.


To my knowledge, the valence-hole concept was first introduced in a review paper on the spectrum of molecular nitrogen by Lofthus and Krupenie~\cite{lofthus1977spectrum}, who conceptually divided the electronic states of N$_2$ into four different groups: $normal$ valence states, valence-hole states, ionic states, and Rydberg states. The authors recognized the specialness of the valence-hole states: their exceptionally large binding energy and dissociation into excited atomic states with one 2p$\leftarrow$2s electron promotion, both of which are distinct from the other $normal$ valence states. Surprisingly, since then, there has not been much discussion of the valence-hole concept until a recent series of papers on the $^1\Pi_u$ and $^3\Pi_u$ states of N$_2$~\cite{sprengers2003extreme,lewis2005predissociation,haverd2005rotational,lewis2008optical,ndome2008sign,lewis2008coupled,vieitez2008complexity, heays2019spin}, the latter of which will be discussed here.

In addition to the appearance of ``globally-linked'' avoided-crossings among multiple adibatic potentials, two additional signatures are identified of valence-hole curve crossings on the global electronic structure~\cite{{jiang2022diabatic},{Borsovszky2021}}. The valence-hole character is transferred to consecutively higher energy adiabatic states following each curve-crossing between the valence-hole state and a normal valence state. This pattern of transfer of valence-hole character can be observed from the electron configuration analysis of an $ab$ $initio$ calculation. A valid global diabatization scheme should reproduce these signature ``jumps" across consecutive adiabatic states as $R\rightarrow\infty$. In addition, for these consecutive electronic states, the vibration-rotation structure deviates significantly from that of a near-rigid-rotor on a Morse-like potential.  As shown in Fig.~\ref{fig:c2_3pig}c, a strongly non-linear vibrational-level ($v$) dependence of the vibration-rotation constants (e.g., vibrational energy spacings, $\Delta G_{v+1/2}$, and the rotational constants) is observed for all four of the lowest $^3\Pi_g$ electronic states of C$_2$. These systematic anomalies in the energy level structure are well reproduced by the valence-hole model. Some partial modeling results for the C$_2$ $^3\Pi_g$ states (reproduced here in Figs.~\ref{fig:c2_3pig}d and \ref{fig:c2_3pig}e) have been previously reported in Ref\citenum{Borsovszky2021}. In this previous work, the predicted predissociation rates of higher-$v$ ($v>12$) levels of the C$_2$ $e^3\Pi_g$ state are used in a photo-dissociation model for the survival of C$_2$ on comets. Figure~\ref{fig:c2_3pig} serves as a formal presentation of the valence-hole modeling results for the C$_2$ $^3\Pi_g$ states. 

Along with the $^3\Pi_g$ states of C$_2$, the $^2\Sigma^+$ states of CN and the $^3\Pi_u$ states of N$_2$ investigated here are among the most extensively studied systems of electronic states. Rotational-level-resolved spectroscopic data are available for multiple vibrational levels of at least three electronic states in the same electronic symmetry manifold. In this work, these high-resolution observations, which show similarly strong non-linear $v$-dependence as observed in the four $^3\Pi_g$ states of C$_2$, are used as inputs to construct an empirical fit model for the CN $^2\Sigma^+$ states and N$_2$ $^3\Pi_u$ states. Similar to the demonstration on the C$_2$ $^3\Pi_g$ states (Figs.~\ref{fig:c2_3pig}d and \ref{fig:c2_3pig}e)~\cite{Borsovszky2021}, the valence-hole model successfully reproduces the highly electronic- and vibrational-state dependent predissociation rates of the N$_2$ $^3\Pi_u$-state levels (Fig.~\ref{fig:n2_3piu_fit}a)~\cite{lewis2008coupled,heays2019spin}. Rydberg$\sim$valence and Rydberg$\sim$Rydberg interactions are successfully incorporated into this fit model for the N$_2$ $^3\Pi_u$ states, for which the four observed electronic states include two valence and two Rydberg states. 

The electronic structures of SiC and Si$_2$ are qualitatively different from that of the isovalent C$_2$ molecule, because of differences in the energy order of the ``analogous" diabatic electronic states~\cite{bruna1980theoretical,peyerimhoff1982potential}. As I will discuss later, the valence-hole state in the $^3\Pi_g$ symmetry manifold of Si$_2$ and the $^3\Pi$ symmetry manifold of SiC both lie higher in energy than some of the $normal$ valence states of the same electronic symmetry. As a result of the higher relative energy of the valence-hole state, a qualitatively different set of curve-crossings between the valence-hole state and the normal valence states exists in Si$_2$ and SiC, which is not present in the C/N/O diatomic molecules. In the two electronic state systems of Si$_2$ and SiC discussed here, the valence-hole state crosses the other valence states at $both$ the inner and outer arms of the valence-hole potential. In the C/N/O diatomic molecules, the valence-hole model has so far only identified curve-crossings on the outer $R$ region of the valence-hole state. The valence-hole electronic structure model for the low-lying Si$_2$ $^3\Pi_g$ and SiC $^3\Pi$ states suggests that the relatively high-lying nature of the valence-hole state in Si$_2$ and SiC is the fundamental reason for the unusual spectroscopic properties of the $L^3\Pi_g$ ($4^3\Pi_g$) state of Si$_2$ and the $C^3\Pi$ state of SiC. The $L^3\Pi_g$ state of Si$_2$, despite being the $fourth$-ranked state in its electronic symmetry block, has the $smallest$ $R_e$ among all $^3\Pi_g$ valence states of Si$_2$~\cite{peyerimhoff1982potential,douglas1955spectrumSi2}. The radiative lifetimes of the $C^3\Pi$ state of SiC are strongly $v$-dependent, with the lifetime decreasing from nearly 3\,$\mu$s at $v=0$ to 500\,ns at $v=6$~\cite{ebben1991c, trinder1993theoretical}.

\section{Methods}
\label{sec:methods}

The diabatic valence-hole model provides a unified treatment of all of the strong interactions between the strongly-bound valence-hole states and numerous other $normal$ valence states. Unlike an $ab$ $initio$ calculation, which ideally treats every possible configuration interaction, only those interactions that directly lead to curve-crossings need to be explicitly treated in the diabatic model here. The effects of configuration interactions from distant states are effectively folded into the diabatic potentials and the electrostatic interaction matrix elements between the crossing diabats. By using analytical functions to model these diabatic potentials with a small number of adjustable parameters and simple assumptions about the $R$-dependence of the interaction matrix elements~\cite{Bob2004}, one constructs an empirical model for the global electronic structure of a molecule. This empirical model can be obtained based on a fit to the $ab$ $initio$ calculation (e.g., adiabatic potentials, $R$-dependence of the valence-hole electron configurations, and non-adiabatic interaction matrix elements) or better yet the available spectroscopic observations (e.g., energy level structure and predissociation linewidths). Pathologically abundant and seemingly intractable multi-configuration interactions, apparent from first-principle calculations, can often be condensed into intuitive two-state problems in the diabatic picture. The construction of the global valence-hole fit models for CN, N$_2$, SiC, and Si$_2$ here follows similar treatments as applied in the previous works on C$_2$~\cite{jiang2022diabatic,Borsovszky2021}. A brief description of the models are provided below. Numerical details are available in Section~S2 of the SI. 

\subsection{$ab$ $initio$ fit models}
\label{sec:adia}

For a diabatic interaction model that involves $n$ electronic states, the adiabatic representation is related to the diabatic one by an $n\times n$ matrix. The diagonal matrix elements are the diabatic potential energies, $E_i^d(R)$, and the off-diagonal matrix elements are the electrostatic interaction between these diabats, $H_{ij}^{el}(R)$ (see Eq.~\ref{eq:He}). The adiabatic potentials are obtained by diagonalizing this $n\times n$ matrix as a function of $R$. These adiabatic potentials from the model are then fitted to the $ab$ $intio$ results (see Section~\ref{sec:ab_initio}). The $R$-dependence of the valence-hole configuration on various adiabatic states from the $ab$ $initio$ calculation (typically at $R\sim R_e$ of the valence-hole state) is also used as inputs to constrain the model.

Morse-like potentials~\cite{jia2012equivalence,hua1990four}, as given in Eq.~\ref{eq:bound}, are used to model all of the bound diabatic states (valence and Rydberg) in this work,
\begin{equation} \label{eq:bound}
E_i^d(R)=T_e+D\left(1-\frac{e^{\beta R_e+h}}{e^{\beta R+h}}\right)^2,
\end{equation}
where $D$ is the dissociation energy for that potential curve relative to its minimum at $T_e$. The dissociated states are modeled by exponential decay functions, $E_i^d(R)=A_re^{-k_rR}$, with the exception of two repulsive potentials of SiC (see Table S3 in the SI). To ensure that the diabatic and adabiatic states converge to the same energy in each dissociation channel, the off-diagonal matrix elements, $H_{ij}^{el}(R)$, must all vanish as $R\rightarrow \infty$. For simplicity, exponential decay functions are used to model the $R$-dependence of these interaction matrix elements, $H_{ij}^{el}(R)=\mathcal{H}_{ij}e^{-s_{ij} R}$. Within a specified electronic state system, the exponential decay rates, $s_{ij}$, are assumed to be the same for most pairs of $H_{ij}^{el}$ (see Tables S1-S4 in the SI). As discussed in Ref~\citenum{jiang2022diabatic}, the specific choice of $R$-dependent form of $H_{ij}^{el}$ is of importance secondary to the use of the correct curve-crossing model. The avoided-crossings patterns are most strongly influenced by the shapes of the diabatic potentials and the magnitudes of $H_{ij}^{el}$ at the $R$-values of their intersections. 

\subsection{Experimental fit models}
\label{sec:adia}

To fit to the experimental observations of the C$_2$ $^3\Pi_g$, CN $^2\Sigma^+$, and N$_2$ $^3\Pi_u$ states, we construct, in each case, an effective Hamiltonian in the diabatic basis. The term energy of each observed vibronic level and its rotational $B$ constant are used as the fit inputs. For the N$_2$ $^3\Pi_u$ states, the observed predissociation linewidths are also used to constrain the model. 

For the Hamiltonian, the diagonal matrix elements are the zeroth-order electronic-vibration-rotation energies of each of the diabats involved in the global model. The off-diagonal electrostatic matrix element between ro-vibrational levels (represented by $\ket{\xi_{v}^d J}$) of two interacting diabatic electronic states (represented by $\ket{\Phi^d}$) are given by 
\begin{equation} \label{eq:He}
H_{i,v_i,J;j,v_j,J}=\bra{\Phi_i^d\xi_{v_i}^d J} H^{el} \ket{\Phi_j^d\xi_{v_j}^d J}=\bra{\xi_{v_i}^d J} H^{el}_{ij}(R) \ket{\xi_{v_j}^d J},
\end{equation}
where $H^{el}_{ij}(R)=\bra{\Phi_i^d} H^{el}\ket{\Phi_j^d}$ is the electronic part of the elecrostatic matrix element.  

Morse-like potentials (Eq.~\ref{eq:bound}) and exponential decay functions are used in the experimental fit models to represent, respectively, bound and dissociated diabatic states. As in the previous works~\cite{jiang2022diabatic,Borsovszky2021}, the ro-vibrational energies of each electronic diabatic state are calculated using the discrete variable representation (DVR) method~\cite{colbert1992novel}. The off-diggonal matrix elements (Eq.~\ref{eq:He}) are obtained by numerical integration of the products of the DVR wavefunctions and $H^{el}_{ij}(R)$. Exponential decay functions with a fixed decay constant ($s_{ij}=1/\textup{\AA}$) are used for the $H^{el}_{ij}(R)$ terms in the fit model for the C$_2$ $^3\Pi_g$ and CN $^2\Sigma^+$ states. A mix of exponentially decaying and $R$-independent $H_{ij}^{el}$ (i.e., $s_{ij}=0/\textup{\AA}$) are used in the fit model for the N$_2$ $^3\Pi_u$ states (see Section~S2 of the SI for details). 




\subsection{Quantum chemical calculation}
\label{sec:ab_initio}

The $ab$ $initio$ potentials for the C$_2$ $^3\Pi_g$ states in Fig.~\ref{fig:c2_3pig}b are obtained using the multi-reference configuration interaction (MRCI) method with the aug-cc-pCVQZ basis, implemented using the MOLPRO program~\cite{werner2020molpro}. The reference space is computed with a full-valence complete active space self-consistence field (CASSCF) calculation of the lowest $^3\Pi_g$ state. The MRCI wavefunctions contain all single and double excitations from the CASSCF reference wavefunction, including excitations from the 1s core. The Davidson correction has been applied to account for the contribution to the correlation energy from the quadruple excitations.

The $ab$ $initio$ calculations for the N$_2$ $^3\Pi_u$, CN $^2\Sigma^+$, SiC $^3\Pi$, and Si$_2$ $^3\Pi_g$ states are obtained using the CASSCF method, implemented with the ORCA quantum chemistry package~\cite{neese2020orca}. The full valence active space includes all of the MOs that arise from the 2s and 2p orbitals of the second-row atom, and the 3s and 3p orbitals of Si (cc-pVTZ basis). These full-valence CASSCF calculations are expected to capture the key configuration interactions among the valence states. The purpose of these calculations is to facilitate an understanding of the global curve-crossing patterns between the valence-hole state and the $normal$ valence states.

\section{$^3\Pi_u$ states of N$_2$}
\label{sec:N2}

\begin{figure}
\includegraphics[width=6.5 in]{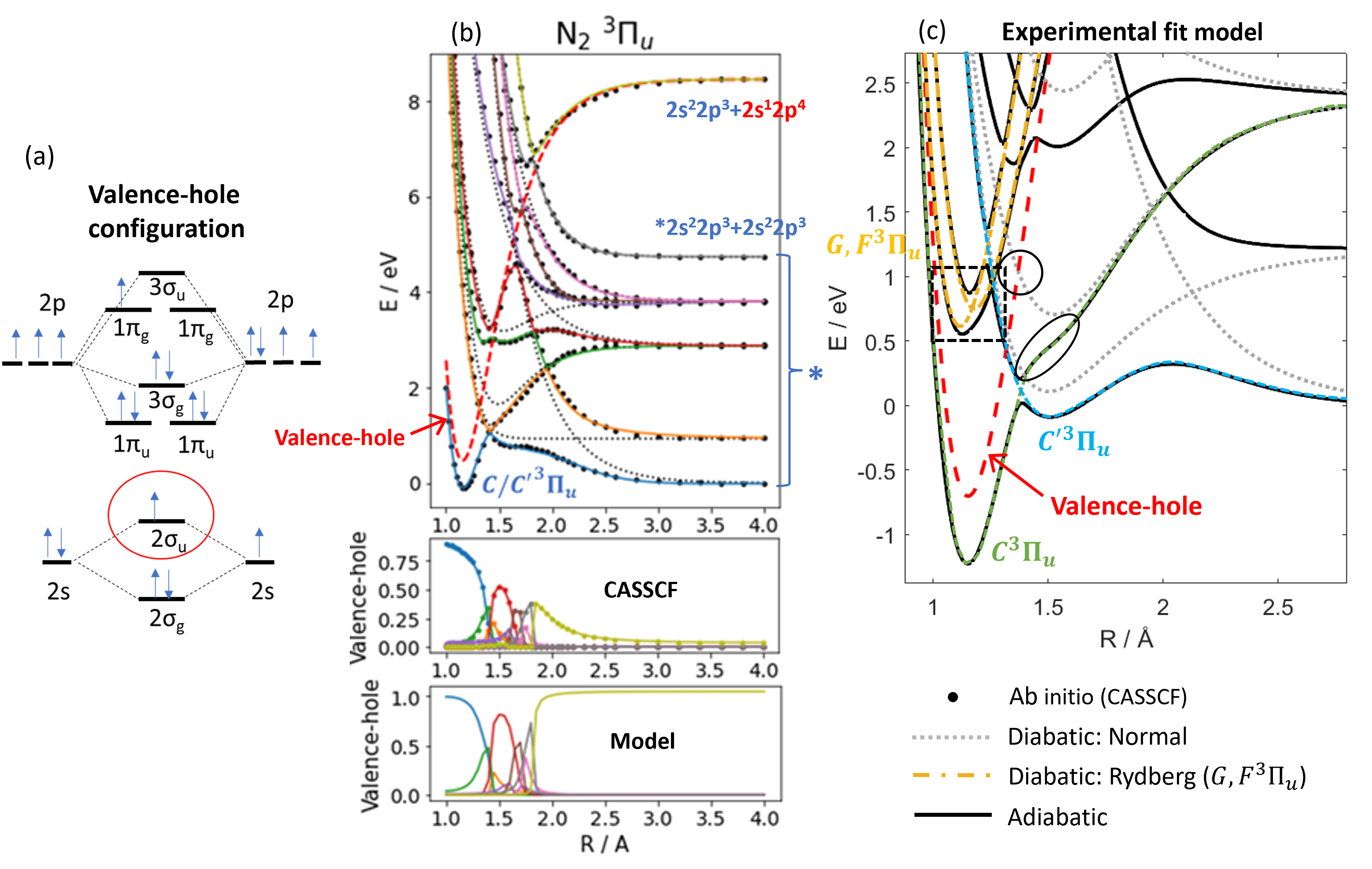}
\caption{Valence-hole model for the $^3\Pi_u$ electronic states of N$_2$. (a) Valence-hole configuration at the $R_e$ of the $C^3\Pi_u$ state. (b) Diabatization scheme for the CASSCF calculation. Top panel: Potential energies from the CASSCF calculation (black dots), and the diabatic and adiabatic potentials from the valence-hole model. Two lower panels: Comparison of the $R$-dependence of the valence-hole character from the CASSCF electron configuration analysis and the valence-hole model. (c) Diabatic and adiabatic potentials derived from the fit to the experimental observations. The diabatic valence-hole state is highlighted using a dashed red line in panels (b) an (c).} 
\label{fig:n2_3piu_calc}
\end{figure}

Two minima exist on the lowest $^3\Pi_u$ adiabatic potential of N$_2$, which are conventionally assigned as the $C^3\Pi_u$ and $C'^3\Pi_u$ states (Fig.~\ref{fig:n2_3piu_calc}c). The $C$ state is dominated by a valence-hole configuration (Fig.~\ref{fig:n2_3piu_calc}a) near $R_e$, and the $C'$ state is composed of $normal$ configurations with a $2\sigma_g^22\sigma_u^2$ valence-core at $R_e$. Carroll and Mulliken~\cite{carroll19653pi} studied the perturbations between the $C$- and $C'$-state ro-vibrational levels, and proposed that the zeroth-order $C$- and $C'$-state potentials intersect at $R \sim1.4$\,\textup{\AA}. This is a weak avoided crossing between the two lowest $^3\Pi_u$ adiabatic states. Based on the predissociation behaviors of the N$_2$ $C$ state~\cite{hori1941pradissoziation,buttenbender1934struktur}, Carroll and Mulliken~\cite{carroll19653pi} further proposed the presence of a barrier at $R\sim2\textup{\AA}$ on the lowest $^3\Pi_u$ adiabatic potential, i.e., at the outer-R region of the $C'$ state (Fig.~\ref{fig:n2_3piu_calc}c). 


A crucial difference exists between the electronic structure of C$_2$ and N$_2$. The valence-hole states in the four electronic symmetry manifolds of C$_2$ investigated earlier~\cite{jiang2022diabatic,Borsovszky2021} are relatively low-lying states, with their electronic state origins $<$5\,eV above that of the ground electronic state. Interactions with the Rydberg states do not need to be directly accounted for in the valence-hole model for these four electronic symmetry species of C$_2$, because even the lowest-energy Rydberg states lie higher in energy than all of the relevant electronic states in the model~\cite{jiang2022diabatic,Borsovszky2021}. In comparison, even the lowest-energy valence-hole state of N$_2$ (which exists in the $^3\Pi_u$ manifold discussed here) lies $>$10 eV above the N$_2$ ground electronic state. To model the energy level structure and predissociation dynamics of the $^3\Pi_u$ states of N$_2$, the Rydberg states and their interactions with the valence-hole and normal valence states must be incorporated into the electronic structure model. Experimentally, the $3p\,\pi_u$ $G^3\Pi_u$ and $3s\,\sigma_g$ $F^3\Pi_u$ states are the observed Rydberg states that converge, respectively, to the $^2\Sigma_g$ and $^2\Pi_u$ electronic states of N$_2^+$.

Before incorporating the Rydberg states into the valence-hole model, I have obtained a full-valence CASSCF calculation for the $^3\Pi_u$ states of N$_2$. The adiabatic potentials and their signature valence-hole character-``jumps'' from the electron configuration analysis (Fig.~\ref{fig:n2_3piu_calc}b) are both well reproduced by the valence-hole model. Note that the contribution from the valence-hole conﬁguration of Fig.~\ref{fig:n2_3piu_calc}a in the total CASSCF wavefunction decreases as $R$ increases. This dilution at large $R$ is due to the increasing number of nearly degenerate valence-hole-type configurations (i.e., with either the $2\sigma_g^22\sigma_u$ or $2\sigma_g^12\sigma_u^2$ valence-core), which inevitably become highly mixed as $R\rightarrow\infty$. After further addition of Rydberg states into diabatization, the valence-hole model reproduces the $R$-matrix calculations for the N$_2$ $^3\Pi_u$ states from Little and Tennyson~\cite{little2013ab}, which simultaneously treat the valence and Rydberg states (see Fig.~S1 in the SI). 

\begin{figure}
\includegraphics[width=6 in]{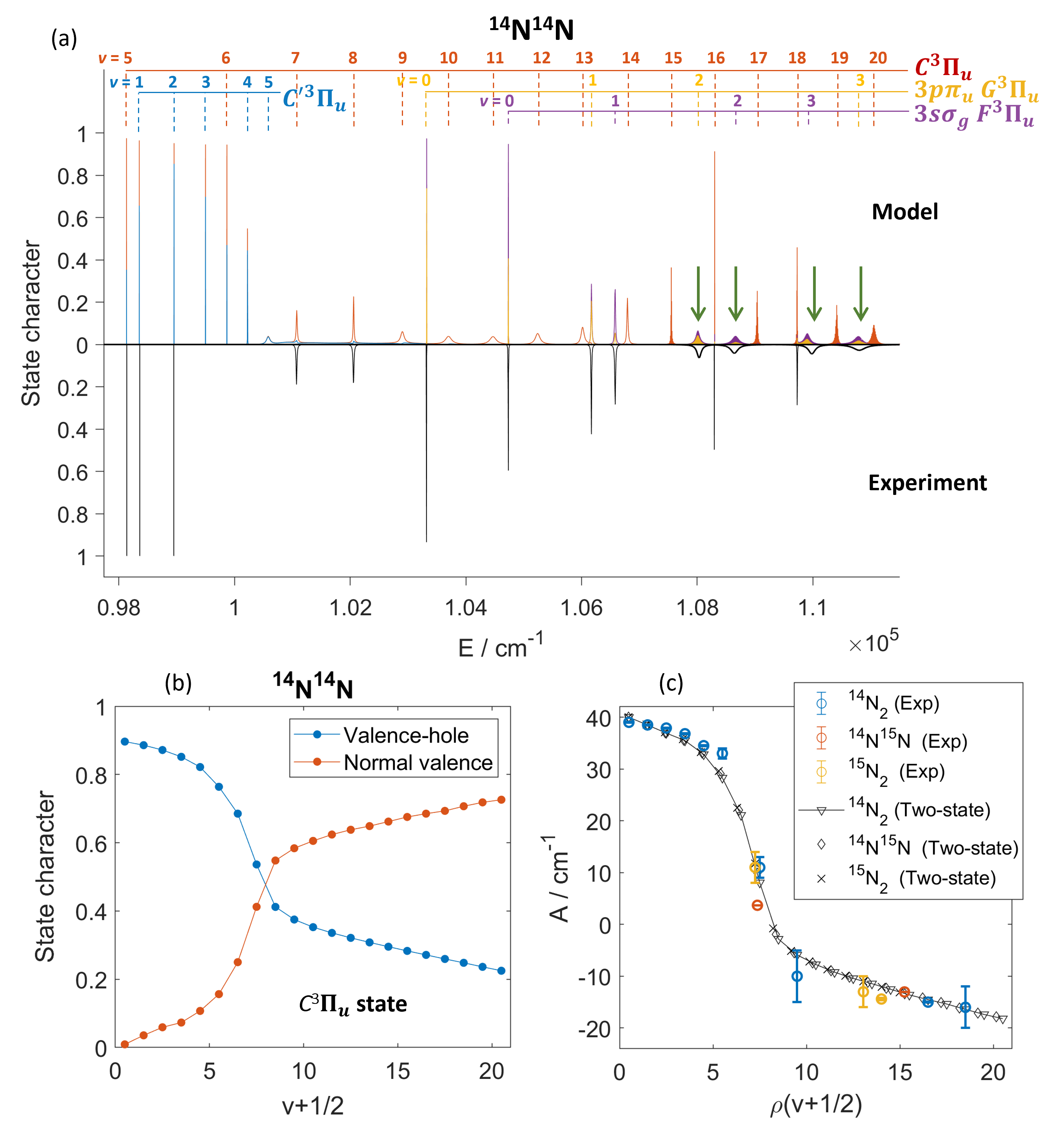}
\caption{Valence-hole model for the $^3\Pi_u$ electronic states of N$_2$, based on the fit to the experiment. (a) Comparison of the model output ($J=1$) and the experimental observations. The experimental bound-state character plot in the lower panel is generated based on the experimental observations summarized in Table I of Ref~\citenum{lewis2008coupled} (vibrational level energy, $T_{exp}$, and predissociation linewidth, $\Gamma_{exp}$). The $\Gamma_{exp}$ values are not available for $v=2$,\,3 of the $G$ and $F$ Rydberg states (indicated by the four arrows in panel a). For these four levels, the $\Gamma_{\textup{CSE}}$ values from the coupled-channel model of Ref~\citenum{lewis2008coupled} are used as the predissociation linewidths. (b) Character of the valence-hole state and one of the normal valence states in different $v$-levels of the $^{14}$N$_2$ $C$ state. The diabatic interaction matrix element responsible for the perturbation between $C$- and $C'$-state levels is set to zero to generate this state character plot. (c) Comparison of the predicted (by a two-state model) and deperturbed experimental spin-orbit $A$ constants~\cite{budo1935dietriplett,ledbetter1976interaction,lewis2008optical,haverd2005rotational,ndome2008sign,heays2019spin} for different $C$-state $v$-levels of the three N$_2$ isotopologues. Mass-reduced vibrational quantum numbers are used for the x-axis.}
\label{fig:n2_3piu_fit}
\end{figure}

Guided by the global diabatization scheme of these $R$-matrix potentials~\cite{little2013ab} and the empirical adiabatic potentials derived from the coupled-channel model of Ref~\citenum{lewis2008coupled}, I have constructed an empirical model for the electronic structure of the N$_2$ $^3\Pi_u$ states, based on the valence-hole concept, using the available experimental inputs from three isotopologues, $^{14}$N$_2$, $^{14}$N$^{15}$N, and $^{15}$N$_2$~\cite{van1994charge,lewis2008optical,sprengers2003extreme,hashimoto2006predissociative,tilford1965high,roux1989high,roux1993high,ledbetter1976interaction,ledbetter1977new,tanaka1961new}. The diabatic potentials and the resulting adiabats from the fit model are shown in Fig.~\ref{fig:n2_3piu_calc}c. In my model, the sudden change of curvature on the conventionally-assigned N$_2$ $C$-state potential (i.e., at the region highlighted by a tilted oval in Fig.~\ref{fig:n2_3piu_calc}c) is caused by a strong interaction between the valence-hole state and a normal valence state. These two diabatic potentials cross each other in the region indicated by a circle in Fig.~\ref{fig:n2_3piu_calc}c. In the model, the barrier in the outer-$R$ region of the $C'$ state is treated as the result of a very strong interaction between a bound and repulsive diabatic state. The inner arm of this bound diabat crosses the outer arms of the valence-hole state and both Rydberg states ($G$ and $F$). This bound state is thus the doorway state that mediates predissociation of the two Rydberg states and the valence $C$ state. In addition to Rydberg$\sim$valence interactions, Rydberg$\sim$Rydberg interaction between the $G$ and $F$ states is significant~\cite{lewis2008coupled,little2013ab}, as is evident from the size of the avoided-crossing that results from the interaction between the two crossing zeroth-order Rydberg potentials in the dashed boxed region of Fig.~\ref{fig:n2_3piu_calc}c.

In Fig.~\ref{fig:n2_3piu_fit}a, the results of the fit model are compared to the experimental observations for the $^{14}$N$_2$ levels above the first dissociation limit. The experimental observations for $^{14}$N$^{15}$N and $^{15}$N$_2$ isotopologues are reproduced with similar quality (see Figs.~S2 and S3 in the SI). For the model, the total bound diabatic state character is shown as a function of energy. The bound-continuum interaction leads to a spread of the bound character into the nearby continuum. As a result, resonances with larger widths have smaller peak heights. The nominal electronic and vibrational state assignments of these resonances are indicated with color-coded tie-lines in Fig.~\ref{fig:n2_3piu_fit}a. Note that the conventional electronic state assignment scheme of N$_2$ (i.e., with the $C$ and $C'$ valence states, and the $G$ and $F$ Rydberg states) is used for this purpose. In Fig.~\ref{fig:n2_3piu_fit}a, the bound state character is further broken down into contributions from the four conventionally-assigned electronic states. This electronic-state-contribution breakdown is illustrated with the same color-code as is applied for the electronic-state assignment with the tie-lines. The state composition analysis indicates that the two Rydberg states are strongly mixed with each other at $v=0-3$, in agreement with previous work by Lewis and co-authors~\cite{lewis2005predissociation,lewis2008coupled,ledbetter1976interaction}. Furthermore, the observed $C$-state $v=5$ level is perturbed by the nearly degenerate $C'$-state $v=1$, which was first modeled by Carroll and Mulliken~\cite{carroll19653pi}.

In the work of Lewis et. al.~\cite{lewis2008coupled}, a coupled-channel  model is used to capture the energy level structure and predissociation linewidths of the N$_2$ $^3\Pi_u$-state levels. The four conventionally-assigned $^3\Pi_u$ electronic states of N$_2$ are treated as the diabatic basis states in this model. Unlike the valence-hole model, the large change of curvature of the $C$-state potential (i.e., at the tilted-oval region in Fig.~\ref{fig:n2_3piu_calc}c) and the barrier in the outer-$R$ region of the $C'$ state are both considered as ``built-in" features of these two diabatic electronic states in Lewis' model. A qualitative two-state interaction argument was proposed in Refs~\citenum{lewis2005predissociation} and \citenum{lewis2008coupled} to explain the ``unusually-shaped'' potential required for the $C$ state, and the marked change in its electron configuration as $R$ increases from $R_e$ (i.e., valence-hole configuration at $R_e$ and $normal$ configuration at larger $R$). 

The valence-hole model not only confirms the chemical intuition of Lewis and co-authors, but also provides a quantitatively accurate, physical interpretation of the observed spectroscopic anomalies for the N$_2$ $^3\Pi_u$ states. In addition to the strongly non-linear $v$-dependence of the $\Delta G_{v+1/2}$ and $B$ constants, the model reproduces the sign reversal of the spin-orbit $A$ constants for the N$_2$ $C$-state vibrational levels (Fig.~\ref{fig:n2_3piu_fit}c), which was first discovered by Ndome et. al~\cite{ndome2008sign} (note that the $C-C'$ interactions have been deperturbed in the determination of the experimental spin-orbit $A$ constants in Fig.~\ref{fig:n2_3piu_fit}c). The $A$ constant for the N$_2$ $C$-state, which is close to +30\,cm$^{-1}$ at $v=6$, suddenly turns 
$negative$ at $v=8$. At $v=18$, the $A$ constant is $-$16 cm$^{-1}$. The valence-hole fit model indicates that, after excluding the diabatic interaction matrix element responsible for the localized $C$$\sim$$C'$ interaction, the resulting ``deperturbed'' $C$-state vibrational levels ($v=0-21$) are well represented by a two-state interaction picture, which accounts for $>$90$\%$ of the total character (Fig.~\ref{fig:n2_3piu_fit}b). Using this simplified two-state model, the observed $v$-dependence of the $A$ constants for all three N$_2$ isotopologues is reproduced if we assume $A=+45$ cm$^{-1}$ for the valence-hole state, and $A=-39$ cm$^{-1}$ for the interacting normal valence state (Fig.~\ref{fig:n2_3piu_fit}c). The valence-hole state of N$_2$ discussed here is a $^3\Pi$ state dominated by a $\sigma\pi$ electron configuration at $R_e$. The $A$ constant for this valence-hole state is expected (and confirmed by the fit) to be of the same $sign$ and similar $magnitude$ as that for the $B^3\Pi_g$ state of N$_2$ (+42 cm$^{-1}$)~\cite{bullock1971molecular}, which is also dominated by a $\sigma\pi$ configuration at $R_e$ ($2\sigma_g^22\sigma_u^21\pi_{u}^43\sigma_g^11\pi_{g}^1$) (see Chapter 3.4.2 of Ref~\citenum{Bob2004}).



\section{$^2\Sigma^+$ states of CN}
\label{sec:CN}

The $^2\Sigma^+$ states of CN are another extensively studied electronic state system with a relatively low energy valence-hole state. Ro-vibrational levels of very high $v$ have been observed for $X^2\Sigma^+$ (up to $v=18$) and $B^2\Sigma^+$ (up to $v=19$) states~\cite{uhler1915structure,jenkins1938mass,douglas1955spectrum,ito1992emission,ram2006fourier,syme2020experimental}, and $v=0-5$ levels for the $E^2\Sigma^+$ have also been reported~\cite{carroll1956spectrum,lutz1970spectrum,syme2020experimental}. While the ro-vibrational structure of the CN $X$ state is reasonably well-behaved, extensive perturbations exist that affect $every$ vibrational level of its $B$ state, which were first recognized by A. E. Douglas nearly 70 years ago~\cite{douglas1955spectrum}. As I will demonstrate here, these systematic anomalies are again caused by the strongly-bound valence-hole state.

\begin{figure}[t]
\includegraphics[width=5 in]{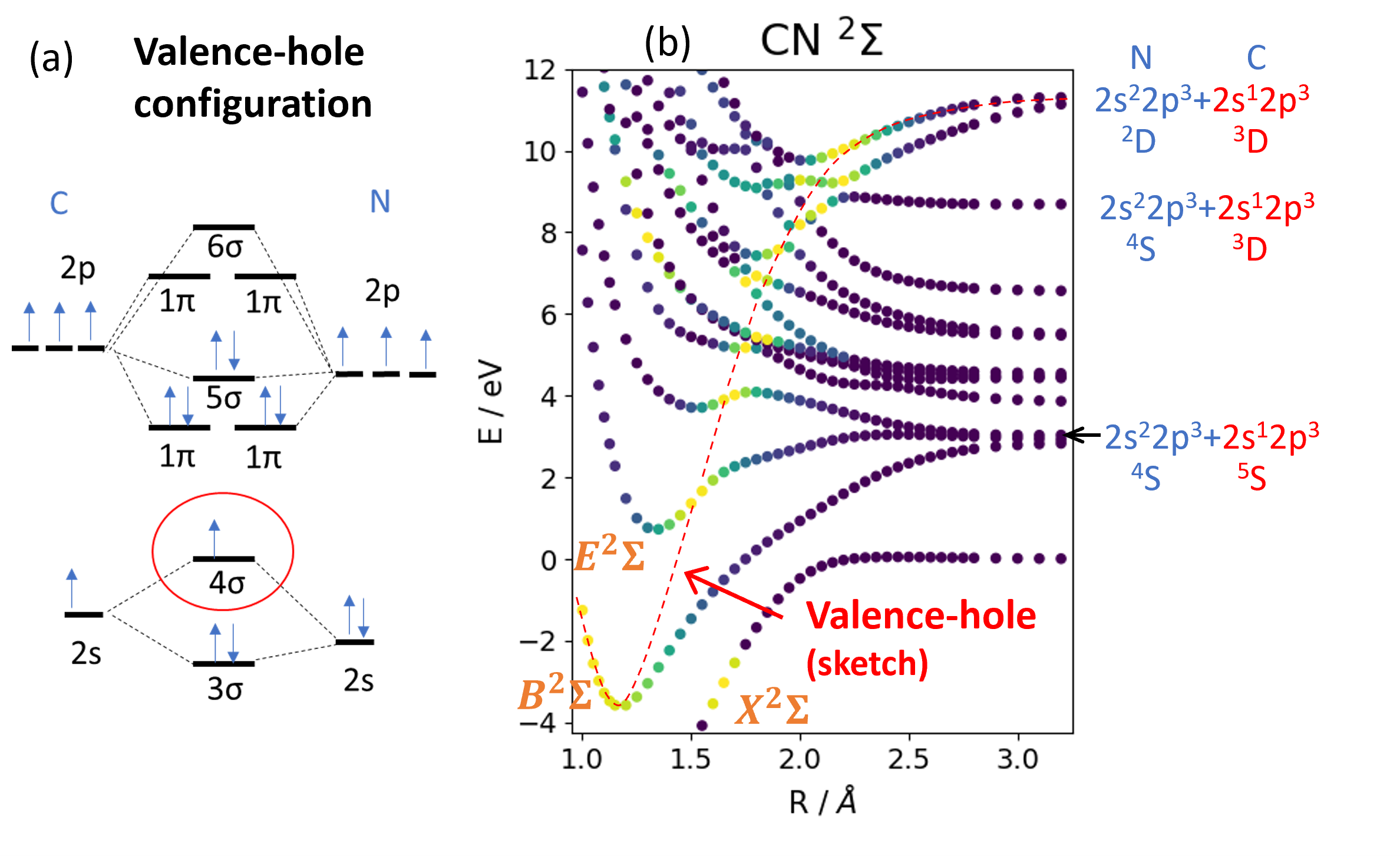}
\caption{Valence-hole state in the $^2\Sigma^+$ electronic symmetry manifold of CN. (a) Valence-hole configuration at the $R_e$ of the $B^2\Sigma^+$ state. (b) Potential energy curves from the CASSCF calculation. A heat map plot, with a yellow (maximum)$-$blue (minimum) gradient, is used to show the $R$-dependence of the character of the valence-hole configuration on each adiabatic state. The appearance of significant valence-hole character at the yellow region of the $X$-state potential in panel (b) is an artifact of the use of the heat map. The valence-hole character accounts for $\sim$0.1$\%$ of the $X$-state wavefunction at this $X$-state maximum region.}
\label{fig:CN_2sigma_calc}
\end{figure}

The valence-hole configuration in the $^2\Sigma^+$ symmetry manifold of CN, as shown in Fig.~\ref{fig:CN_2sigma_calc}a, is excepted be strongly bound, considering its nominal bond order of 3.5. This valence-hole configuration is the dominant configuration at the $R_e$ of the CN $B$ state ($\sim$80$\%$ of the total state character)~\cite{schaefer1971electronic}. While a global diabatization for the numerous $^2\Sigma^+$ potentials in Fig.~\ref{fig:CN_2sigma_calc}b has not been implemented, the global signature of valence-hole curve crossings is clear from the results of the electron configuration analysis of the CASSCF calculation. In Fig.~\ref{fig:CN_2sigma_calc}b, the $R$-region with the largest valence-hole character on a given adiabatic potential is highlighted in yellow, while that with the smallest valence-hole contribution is in dark blue. The use of this heat map plot allows one to qualitatively trace the diabatic valence-hole potential by following the character of a single valence-hole configuration (Fig.~\ref{fig:CN_2sigma_calc}a), even though the contribution of this single configuration in the total CASSCF wavefunction decreases as $R\rightarrow\infty$. As is evident from Fig.~\ref{fig:CN_2sigma_calc}b, as $R$ increases, the valence-hole character jumps from the $B$ state to consecutively higher-energy adiabatic states, before dissociating into a very high energy channel. 

\begin{figure}[t]
\includegraphics[width=6.2 in]{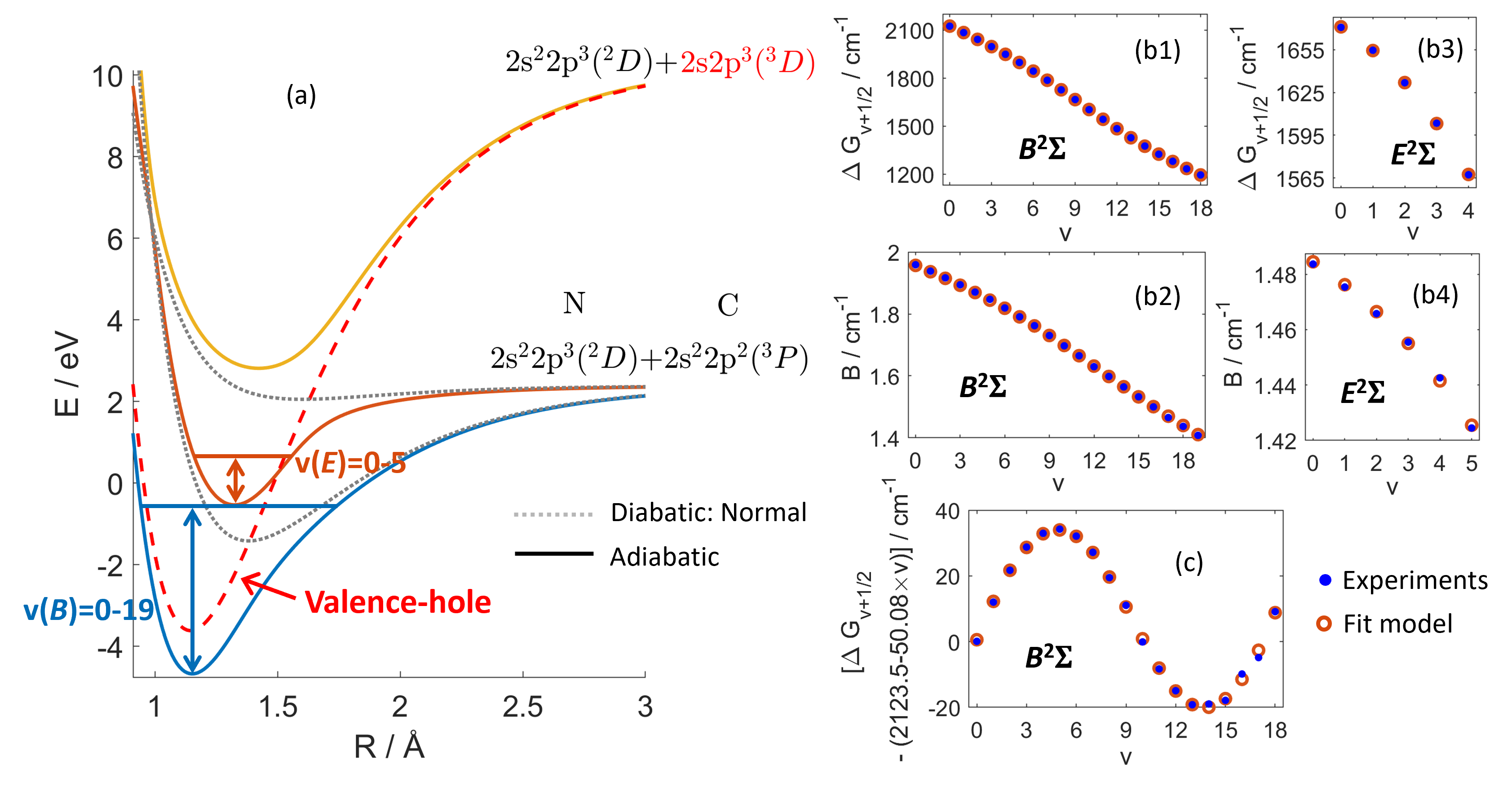}
\caption{Three-state valence-hole model for the $^2\Sigma^+$ electronic states of CN, based on the fit to the experiment. (a) Diabatic potentials and the corresponding adiabatic potentials derived from the fit to the experiment. (b1)-(b4) Comparison of the observed $\Delta G_{v+1/2}$ values for the $B$ and $E$ states with the fit model output. The vibronic term energies and the rotational $B$ constants are reproduced, respectively, with a mean absolute error of 0.33 cm$^{-1}$  and 0.0013 cm$^{-1}$. (c) Oscillation in the $\Delta G_{v+1/2}$ values for the $B$ state from the experiment and the fit model.}
\label{fig:CN_2sigma_fit}
\end{figure}

To further demonstrate the existence of the global valence-hole curve-crossings inferred from Fig.~\ref{fig:CN_2sigma_calc}b, I have constructed a three-state interaction fit model to reproduce the observed vibration-rotation structure of the CN $B$ and $E$ states. In this three-state model (Fig.~\ref{fig:CN_2sigma_fit}), the diabatic valence-hole state dissociates into the C (2s2p$^3$, $^3D$) +N (2s$^2$2p$^3$, $^2D$) fragment limit. The $X$ state, which lies $>$3\,eV lower than the $B$ state, is assumed to be unaffected by the valence-hole state, and is excluded from the fit model.

As is evident from Fig.~\ref{fig:CN_2sigma_calc}b, the assumed diabatic dissociation limit for the valence-hole state is $not$ the lowest energy dissociation channel with an excited 2s2p$^3$ carbon. Based on the $R$-dependence of the valence-hole character inferred from Fig.~\ref{fig:CN_2sigma_calc}b, it is clear that neither C (2s2p$^3$, $^5S$) +N (2s$^2$2p$^3$, $^4S$) nor C (2s2p$^3$, $^3D$) +N (2s$^2$2p$^3$, $^4S$) is the correct dissociation limit for the valence-hole state. Furthermore, compared to the alternative assumption of the C (2s2p$^3$, $^5S$) +N (2s$^2$2p$^3$, $^4S$) limit, the residuals of the three-state fit model are significantly improved (by $\sim$10$\times$), when the high-energy C (2s2p$^3$, $^3D$) +N (2s$^2$2p$^3$, $^2D$) channel is taken as the dissociation limit for the valence-hole state. 


The three-state fit model reproduces the vibrational level energies and rotational $B$ constants for both the $B$ and $E$ states (Fig.~\ref{fig:CN_2sigma_fit}b1-\ref{fig:CN_2sigma_fit}b4). A. E. Douglas had noted that the $\Delta G_{v+1/2}$ values for the $B$ state have an underlying oscillatory pattern, which becomes obvious after subtraction of an overall linear trend (Fig.~\ref{fig:CN_2sigma_fit}c)~\cite{douglas1955spectrum}. In my model, these systematic perturbations in the $B$-state vibrational energy structure are caused by curve-crossings between the valence-hole state and $normal$ valence states. On its way to dissociation, the valence-hole state has again left a trail of breadcrumbs of insights into the global electronic structure through perturbations in the molecular spectra.

\section{S\lowercase{i}C \lowercase{and} S\lowercase{i}$_2$}
\label{sec:Si}

The diabatic valence-hole model provides a unified picture of the global electronic structure of molecules beyond the second-row diatomic species. Here, the $^3\Pi_g$ states of Si$_2$ and the $^3\Pi$ states of SiC are chosen for demonstration, motivated by the fact that the ro-vibrational levels of at least one electronic state have been observed in these two electronic state systems ($L^3\Pi_g$ for Si$_2$; $C^3\Pi$ and $X^3\Pi$ for SiC). Given the large cosmic abundance of Si (the eighth most abundant element, with abundance similar to that of N)~\cite{anders1989abundances,lodders2003solar}, Si$_2$ and SiC are both of fundamental importance in astrophysical environments. 

\begin{figure}
\includegraphics[width=6.5 in]{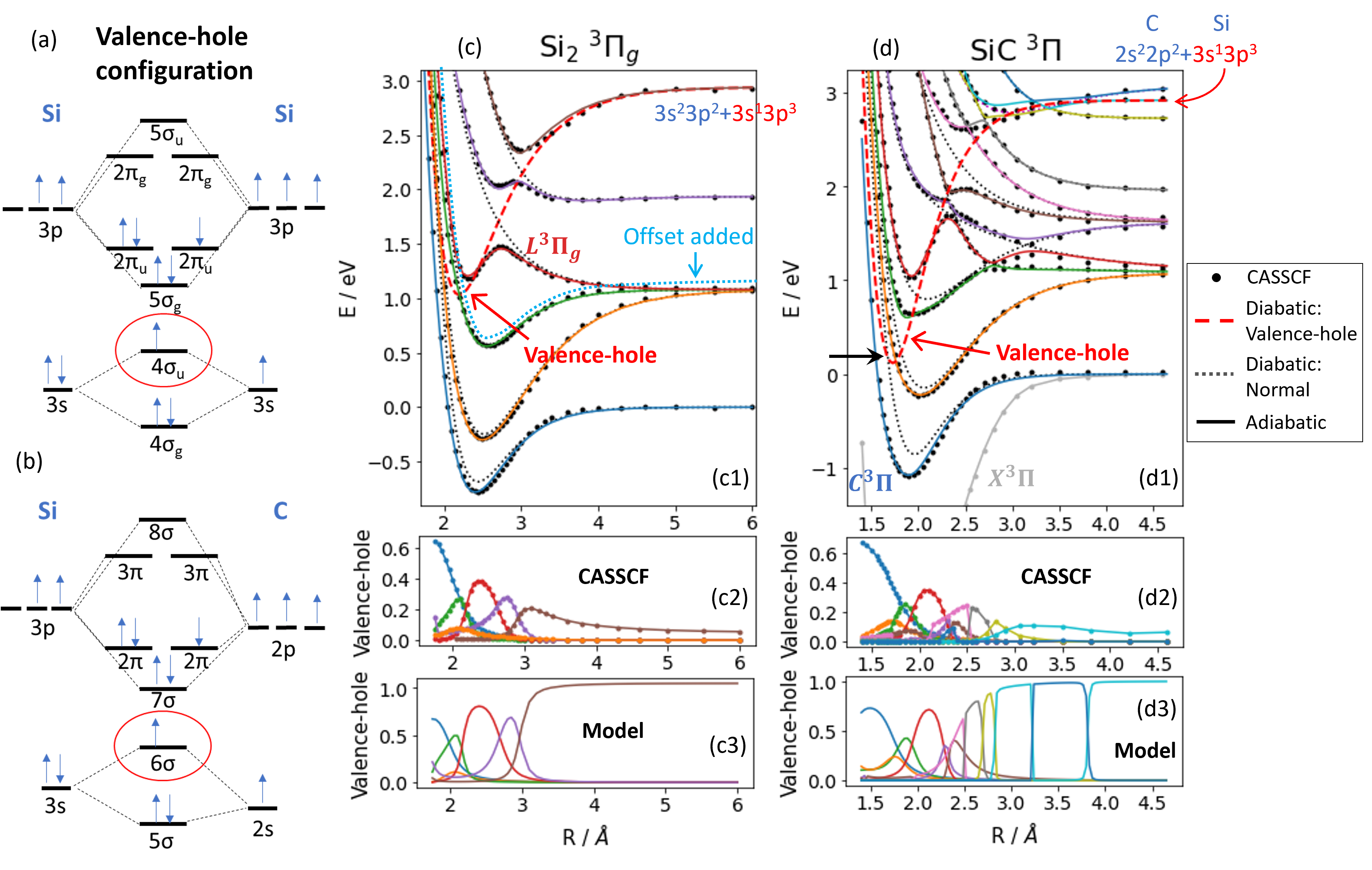}
\caption{Valence-hole model for the $^3\Pi_g$ states of Si$_2$ and the $^3\Pi$ states of SiC. (a) and (b) Relevant valence-hole configurations for Si$_2$ and SiC. (c) and (d) Diabatization scheme for the CASSCF calculation for the Si$_2$ $^3\Pi_g$ states and the SiC $^3\Pi$ states. Top panel: Potential energies from the CASSCF calculation, and the diabatic and adiabatic potentials from the valence-hole model. Two lower panels: Comparison of the $R$-dependence of the valence-hole character from the CASSCF electron configuration analysis and the valence-hole model. To highlight the third normal diabatic state in panel (c1), a small vertical offset is added to its diabatic potential. Similar to the treatment of the $X^2\Sigma^+$ state of CN, the $X^3\Pi$ state of SiC is excluded from the valence-hole diabatization for the SiC $^3\Pi$ states, because the low-lying $X$ state is not expected to be significantly affected by the valence-hole state.}
\label{fig:sic_si2_cal}
\end{figure}

The valence-hole model suggests that the $^3\Pi_g$ valence-hole state of Si$_2$ (Fig.~\ref{fig:sic_si2_cal}c) and the $^3\Pi$ valence-hole state of SiC (Fig.~\ref{fig:sic_si2_cal}d) both lie higher in energy at their respective $R_e$ than many $normal$ valence states in the same electronic symmetry manifold. In comparison, the valence-hole state is the lowest energy diabatic state in all four electronic symmetry manifolds ($^1\Pi_g$, $^3\Pi_g$, $^1\Sigma_g^+$, and $^3\Sigma_g^+$) of C$_2$ investigated earlier~\cite{jiang2022diabatic,Borsovszky2021}. As a result of this different energy order, the curve-crossing patterns are qualitatively different among the Si$_2$ $^3\Pi_g$ and the SiC $^3\Pi$ states. In C$_2$, N$_2$, and CN, curve-crossings between the valence-hole and normal valence states occur on the outer arm of the valence-hole potential (see Figs.~\ref{fig:c2_3pig}, ~\ref{fig:n2_3piu_calc}, and~\ref{fig:CN_2sigma_fit}). For Si$_2$ and SiC (Fig.~\ref{fig:sic_si2_cal}), the valence-hole state crosses other normal valence states on $both$ the inner and outer arms of the valence-hole potential. 


The valence-hole diabatization scheme proposed for Si$_2$ and SiC in Fig.~\ref{fig:sic_si2_cal} is supported by the results of the electron configuration analysis from the CASSCF calculation. As is evident by comparing the bottom two panels in Figs.~\ref{fig:sic_si2_cal}c, the valence-hole model reproduces the $R$-dependence of the valence-hole character on all six $^3\Pi_g$ states of Si$_2$ for the entire $R$-range. As a result of the loss of the $g/u$ symmetry, the $^3\Pi$ electronic state density is approximately doubled for SiC compared to Si$_2$. The higher incidence of curve-crossings between the valence-hole and normal states in the $^3\Pi$ symmetry manifold of SiC leads to transfer of the valence-hole character across nearly 10 states in a small energy ($\sim$4\,eV) and $R$ ($1.6<R<3$ \textup{\AA}) window. This complicated and cluttered pattern of transfer of valence-hole character (Fig.~\ref{fig:sic_si2_cal}d3) reproduces the result of the CASSCF electron configuration analysis at $R<3$\textup{\AA} (Fig.~\ref{fig:sic_si2_cal}d2). However, the model yields two curve-crossings in the larger $R$ region, which are absent from the CASSCF result. I find that these two additional crossings are unavoidable in a fit model that aims to reproduce the curve-crossing dynamics at the smaller $R$. The failure of the model at $R>3$\textup{\AA} is caused by the incorrect long-range behavior of the analytical Morse-like potentials~\cite{jia2012equivalence,hua1990four} used to model the bound diabatic states. 


\begin{figure}[t]
\includegraphics[width=6.5 in]{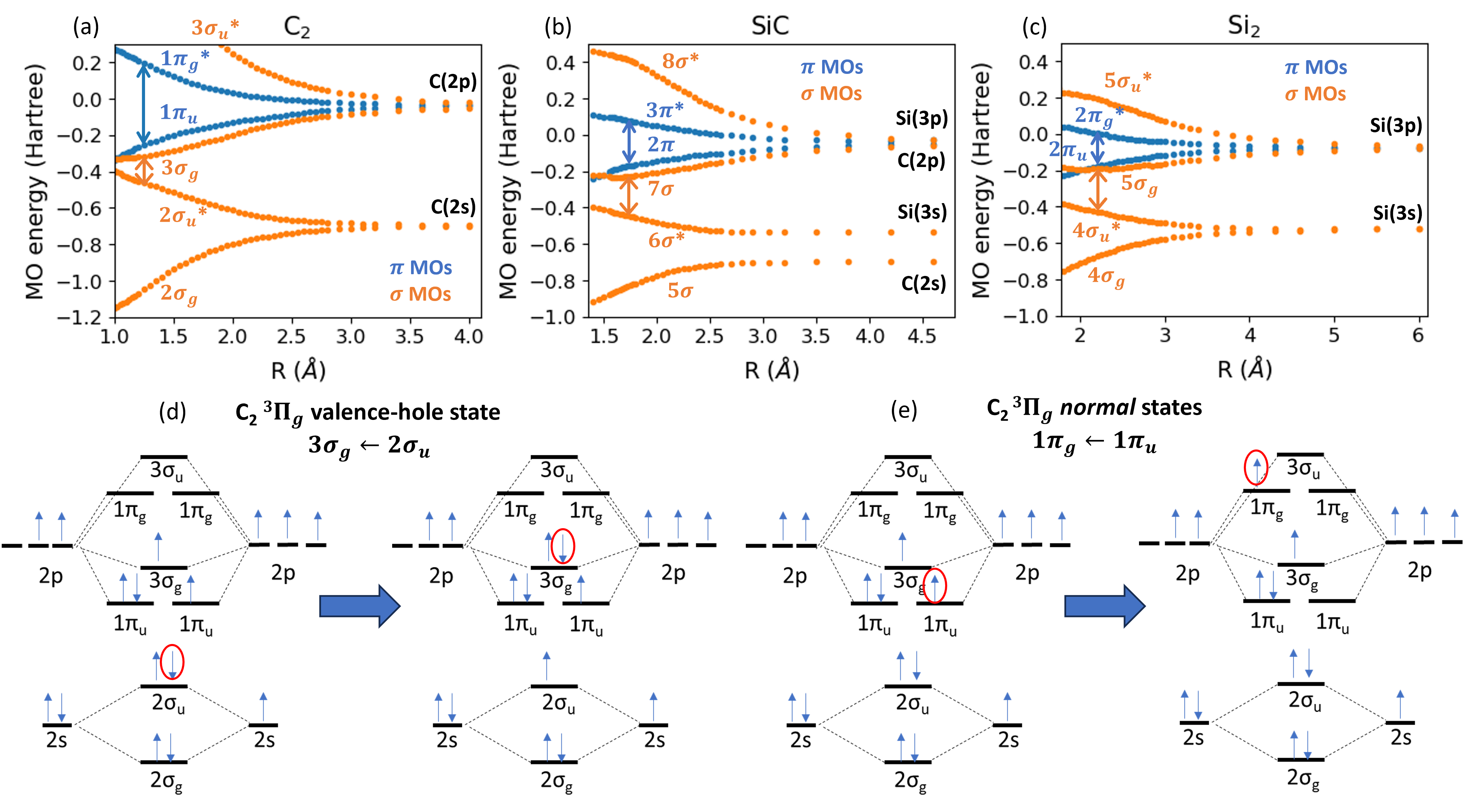}
\caption{Different energy order between the valence-hole and $normal$ valence states in C$_2$, SiC, and Si$_2$. The MO energies from the CASSCF calculations are shown for the three molecules in panels (a)-(c), where the asterisks are used to indicate the anti-bonding MOs and the double-headed arrows to illustrate the energy separations of selected pairs of MOs near the $R_e$ of the relevant valence-hole state. (d) Electron promotion from the lowest $\Pi_u$-symmetry electron configuration of C$_2$ to generate the valence-hole configuration. (e) Electron promotion from the lowest $\Pi_u$-symmetry electron configuration of C$_2$ to generate a $normal$ electron configuration.}
\label{fig:MO_comparison}
\end{figure}

The different energy order between the valence-hole and $normal$ valence states in Si$_2$ and SiC discussed here relative to that in the isovalent C$_2$ molecule is a consequence of the qualitative change in the relative stability of various bonding and anti-bonding MOs. As is evident from Figs.~\ref{fig:MO_comparison}a-\ref{fig:MO_comparison}c, the energy splittings between the anti-bonding and bonding MOs, n$\sigma^*$$-$n$\sigma$ and n$\pi^*$$-$n$\pi$, become progressively smaller from C$_2$ to SiC to Si$_2$ at the small $R$ region near the $R_e$ of the corresponding valence-hole state ($R_e\sim1.25$\,\textup{\AA}, 1.75\,\textup{\AA}, and 2.25\,\textup{\AA}, respectively, for C$_2$, SiC, and Si$_2$). Using the lowest-energy $\Pi$-symmetry electron configuration as the reference (e.g., $2\sigma_g^22\sigma_u^21\pi_{u}^33\sigma_g^1$ for C$_2$, see Figs.~\ref{fig:MO_comparison}d and \ref{fig:MO_comparison}e), the low-lying $normal$ valence states derive from an n$\pi^*$$\leftarrow$n$\pi$ electron promotion for all three electronic state systems discussed here, and the valence-hole state results from an (n+1)$\sigma$$\leftarrow$n$\sigma^*$ electron promotion. In C$_2$, for which the $3\sigma_g-2\sigma_u^*$ energy separation is much smaller than that of $1\pi_g^*-1\pi_u$ (Fig.~\ref{fig:MO_comparison}a), the valence-hole state is the lowest $^3\Pi_g$ state near its $R_e$. In Si$_2$ and SiC, one or more of the $normal$ valence states are lower in energy than the valence-hole state, because of a combination of reduced energy separation for n$\pi^*$$-$n$\pi$ and an increase for (n+1)$\sigma$$\leftarrow$n$\sigma^*$ (Fig.~\ref{fig:MO_comparison}b and \ref{fig:MO_comparison}c).

The relatively higher energy of the valence-hole states in Si$_2$ and SiC leads to unusual spectroscopic properties for the Si$_2$ $L^3\Pi_g$ state and the SiC $C^3\Pi$ state, which are not observed in any of the electronic state system of the C/N/O diatomic species with a lower lying valence-hole state. Here, I will discuss and interpret these peculiar spectroscopic features based on the valence-hole diabatization schemes proposed in Fig.~\ref{fig:sic_si2_cal}. 

The $L^3\Pi_g$ state of Si$_2$ was first assigned by A. E. Douglas from the observation of a $^3\Pi_g$$-$$^3\Pi_u$ band (assigned as $L^3\Pi_g-D^3\Pi_u$)~\cite{douglas1955spectrumSi2}. Only the $v=0$ level of the $L$ state was observed. The higher $v$ levels were believed to suffer from strong predissociation. The multi-reference double-excitation configuration interaction (MRD-CI) calculation from Peyerimhoff and Buenker~\cite{peyerimhoff1982potential} was the first to confirm the electronic state assignment of Douglas. Based on its observed energy and $R_e$, Peyerimhoff and Buenker assigned the $L^3\Pi_g$ state to the fourth $^3\Pi_g$ state of Si$_2$. The maximum $R$ value reported for the $4^3\Pi_g$ state is $\sim$2.4\,\textup{\AA} from this MRD-CI calculation. As a result, the outer barrier on the $4^3\Pi_g$ ($L$) state (at $R\sim2.7$\,\textup{\AA}) predicted by the CASSCF calculation (Fig.~\ref{fig:sic_si2_cal}c1) is absent from the MRD-CI calculation. According to both my CASSCF and earlier MRC-CI results~\cite{peyerimhoff1982potential}, the inner well of the $L^3\Pi_g$ state has a smaller $R_e$ than all three of the lowest $^3\Pi_g$ adiabatic states. This is an unusual situation for valence states. The lowest energy state in a given electronic symmetry manifold is usually expected to have the smallest $R_e$. 

The presence of a relatively high-lying valence-hole state with a small $R_e$ leads to violation of this conventional wisdom. The valence-hole diabatization scheme for the Si$_2$ $^3\Pi_g$ states reproduces the small $R_e$ for the inner well of the $L^3\Pi_g$ state. According to the model (Fig.~\ref{fig:sic_si2_cal}c), the $L$-state inner well is formed as a result of the curve-crossing between the valence-hole state (dashed red) and the third normal state (dotted cyan). The $R_e$ for this inner well is strongly influenced by the small $R_e$ of the valence-hole state. The CASSCF calculation indicates that the absence of observations of higher $v$ levels of the $L^3\Pi_g$ state is most likely a consequence of the shallowless of the $L$-state inner well. In my model, the outer barrier on the $L$-state potential is caused by the strong interaction between the valence-hole state and a repulsive state, which cross each other at $R\sim2.7$\,\textup{\AA}.

The first spectroscopic observation of gas-phase SiC was made by Bernath et. al., who observed the 0-0 band of the $d^1\Sigma^+-b^1\Pi$ electronic transition of SiC~\cite{bernath1988theoretical}. The $C^3\Pi-X^3\Pi$ transitions were observed shortly after this initial experiment~\cite{ebben1991c,butenhoff1991c}. With laser induced fluorescence detection, Ebben et. al. observed a strong $v$-dependence of the radiative lifetimes of the SiC $C^3\Pi$ state, which decrease from nearly 3\,$\mu$s at $v=0$ to 0.5\,$\mu$s at $v=6$~\cite{ebben1991c}. Based on this $v$-dependence of the radiative lifetimes and the intensity patterns in the dispersed fluorescence spectra from various $C$-state vibrational levels, Ebben et. al. proposed that the $C^3\Pi-X^3\Pi$ electronic transition dipole moment must decrease rapidly as $R$ increases, and approach zero near the outer turning point of the $C$-state $v=0$ vibrational wavefunction. A CASSCF calculation~\cite{trinder1993theoretical} confirmed this proposed strong $R$-dependence of the $C^3\Pi-X^3\Pi$ transition dipole moment. The rapid decrease of the transition dipole moment as a function of $R$ is caused by the rapidly decreasing valence-hole character in the SiC $C^3\Pi$ state. As can be seen from Fig.~\ref{fig:sic_si2_cal}d2, the SiC $C$ state is dominated by the valence-hole character at the small $R$ region ($R\lesssim
$1.7\,\textup{\AA}). However, at $R_e$ ($\sim$1.9\,\textup{\AA}), the $C$ state has transferred most of its valence-hole character into higher energy $^3\Pi$ states.

In my model, this rapid and ``premature" loss of the valence-hole character from the SiC $C^3\Pi$ state is caused by the new type of curve-crossing (as previously discussed with Si$_2$), which is not present in the low-lying electronic structure of C$_2$ and other second-row diatomic molecules. Because of the relatively high energy of the valence-hole state and its smallest $R_e$ among the $^3\Pi$ states of SiC, the ``first" curve-crossing (i.e., as $R$ increases) between the valence-hole state and a normal valence state (in the region indicated by the horizontal arrow in Fig.~\ref{fig:sic_si2_cal}d1) occurs at the $inner$ potential arms of the two interacting diabatic states. As a result of the steep slopes at the inner $R$ region, the two diabatic potentials appear to seamlessly blend into each other. The two resulting adiabats ($C^3\Pi$ and $3^3\Pi$ in Fig.~\ref{fig:sic_si2_cal}d1) are thus seemingly well-behaved throughout the entire range of $R$. The innocuous appearances of the adibatic potentials belie the extremely complicated, global-scale, multiple curve-crossings, which are manifest as unusual spectroscopic patterns in the molecular spectra.

\section{Conclusion}
\label{sec:conclusion}


The presence of a valence-hole diabatic state leads to several classes of large, systematic disruptions to the molecular electronic structure. The valence-hole concept frames the ``big picture" of the global curve-crossing dynamics between the valence-hole state and numerous $normal$ valence states. On its way to dissociation into a high-energy separated-atom limit with one np$\leftarrow$ns (n=2 or 3) promoted atomic state, the valence-hole state interacts strongly with these other valence states via interaction matrix elements often as large as 1\,eV. In the adiabatic representation, these successive electrostatic interactions result in an interconnected network of avoid-crossings among $all$ of the low-lying adiabatic states. In my model, the ``wiggles" in the adiabatic potentials are then treated $locally$ on an $ad$ $hoc$ basis as curve-crossings between $normal$ valence states. 

The global diabatization scheme reported here is a $unified$ treatment of multiple heavily perturbed electronic symmetry manifolds of C$_2$, N$_2$, CN, Si$_2$, and SiC, in which one low-lying diabatic valence-hole state exists. This compact diabatic representation of the electronic structure lends insight into the unusual energy level structure, spectroscopic properties (e.g., $R_e$ and radiative lifetimes), and unimolecular dynamics (e.g., predissociation) for these electronically excited states. Some of these electronic state systems, such as the $^3\Pi_g$ states of C$_2$, the $^3\Pi_u$ states of N$_2$, and the $^2\Sigma^+$ states of CN, have been extensively characterized experimentally and theoretically. However, the understanding of these classic electronic states has been surprisingly incomplete, because of the neglect of the existence of the strongly-bound valence-hole state. The presence of systematic perturbations in their energy level structure should have been obvious, even from pre-1960 spectroscopic observations. However, prior to the development of the valence-hole diabatic interaction model, no formal framework had been proposed to account for these anomalies, with the exception of the analysis from Ballik and Ramsay of the C$_2$ $C^1\Pi_g$ state~\cite{Ballik1963}, which inspired the study of the electronic structure of C$_2$, as detailed in the previous work by the author and co-workers~\cite{jiang2022diabatic}. 

Considering the often profound impact of valence-hole states on the energy level structure and unimolecular dynamics of electronically excited states, the long-neglected concept of valence-hole states should become an integral part of our intuitive picture of electronic structure, along with other familiar concepts such as Rydberg states and ionic states. A clear distinction is fully merited between valence-hole states and other types of valence states, because of the qualitative difference in their chemical binding energies and dissociation products. This recognition of the importance of the valence-hole concept has allowed us to answer questions about the molecules that we did not even know how to ask.

\begin{acknowledgement}

The author would like to thank Prof. Robert W. Field (MIT) for his insight and encouragement throughput this project. This work was carried out at the Lawrence Livermore National Laboratory under the auspices of the U.S. Department of Energy by Lawrence Livermore National Laboratory under Contract DE-AC52-07NA27344.

\end{acknowledgement}

\begin{suppinfo}
Supplemental Information: Supplemental figures (Section S1); Numerical details of various global diabatic models (Section S2). Results of the $ab$ $initio$ calculation are included in a separate Excel document.
\end{suppinfo}

\newpage
\bibliography{vh_JPCA}

\end{document}




\newpage

\section{Supplemental figure}
\label{sec:figures}

\begin{figure}
\center
\includegraphics[width=4.5 in]{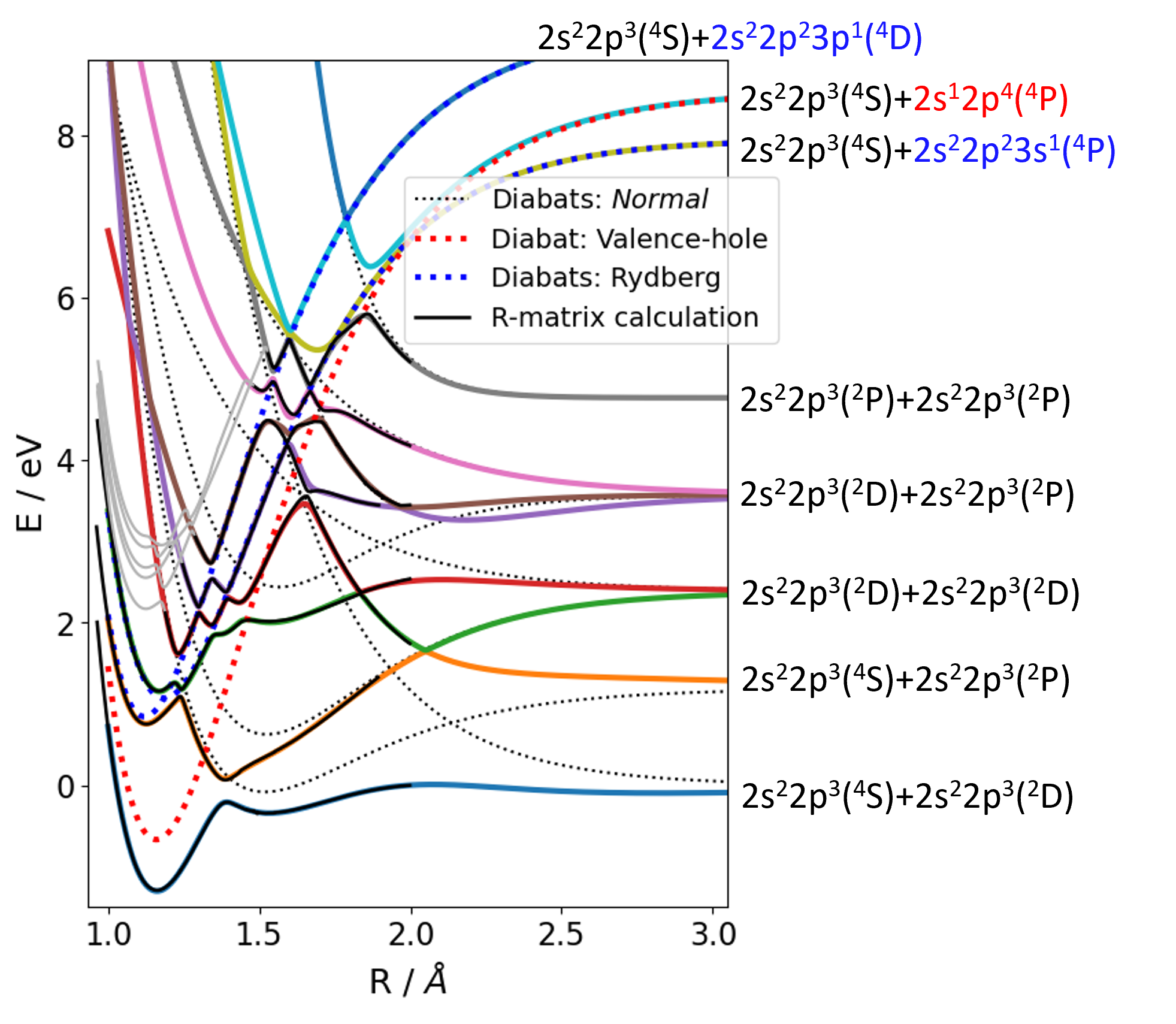}
\caption{Valence-hole global diabatization scheme for the $R$-matrix potentials of the $^3\Pi_u$ electronic states of N$_2$ from Ref~\citenum{little2013ab}. Higher Rydberg states (solid gray curves) are not included in the fit model.}
\label{fig:R_matrix_fit}
\end{figure}

The global diabatization scheme for the $R$-matrix potentials of the $^3\Pi_u$ electronic states of N$_2$ from Ref~\citenum{little2013ab} is shown in Fig.~\ref{fig:R_matrix_fit}. The full-valence CASSCF calculation for the N$_2$ $^3\Pi_u$ states (Fig.~2b of the main text) and the empirical adiabatic potentials from Ref~\citenum{lewis2008coupled} are used to constrain the shape of the adiabatic potentials from the valence-hole fit model (indicated by the solid colored lines in Fig.~\ref{fig:R_matrix_fit}) at the outer $R$ region ($R\gtrsim2$\,\textup{\AA}), where the $R$-matrix calculation is not available. Using the MO correlation diagram by Mulliken~\cite{mulliken1966rydberg}, the zeroth-order $3p\,\pi_u$ $^3\Pi_u$ and $3s\,\sigma_g$ $^3\Pi_u$ Rydberg states are assumed to dissociate, respectively, into the lowest-energy separated-atom limit (allowed by the electronic-state symmetry) with one nitrogen atom in the 3p and 3s Rydberg configuration, i.e., 2s$^2$2p$^3$($^4$S)+2s$^2$2p$^2$3p($^4$D) for the former and 2s$^2$2p$^3$($^4$S)+2s$^2$2p$^2$3s($^4$P) for the latter.

\begin{figure}
\includegraphics[width=5.5 in]{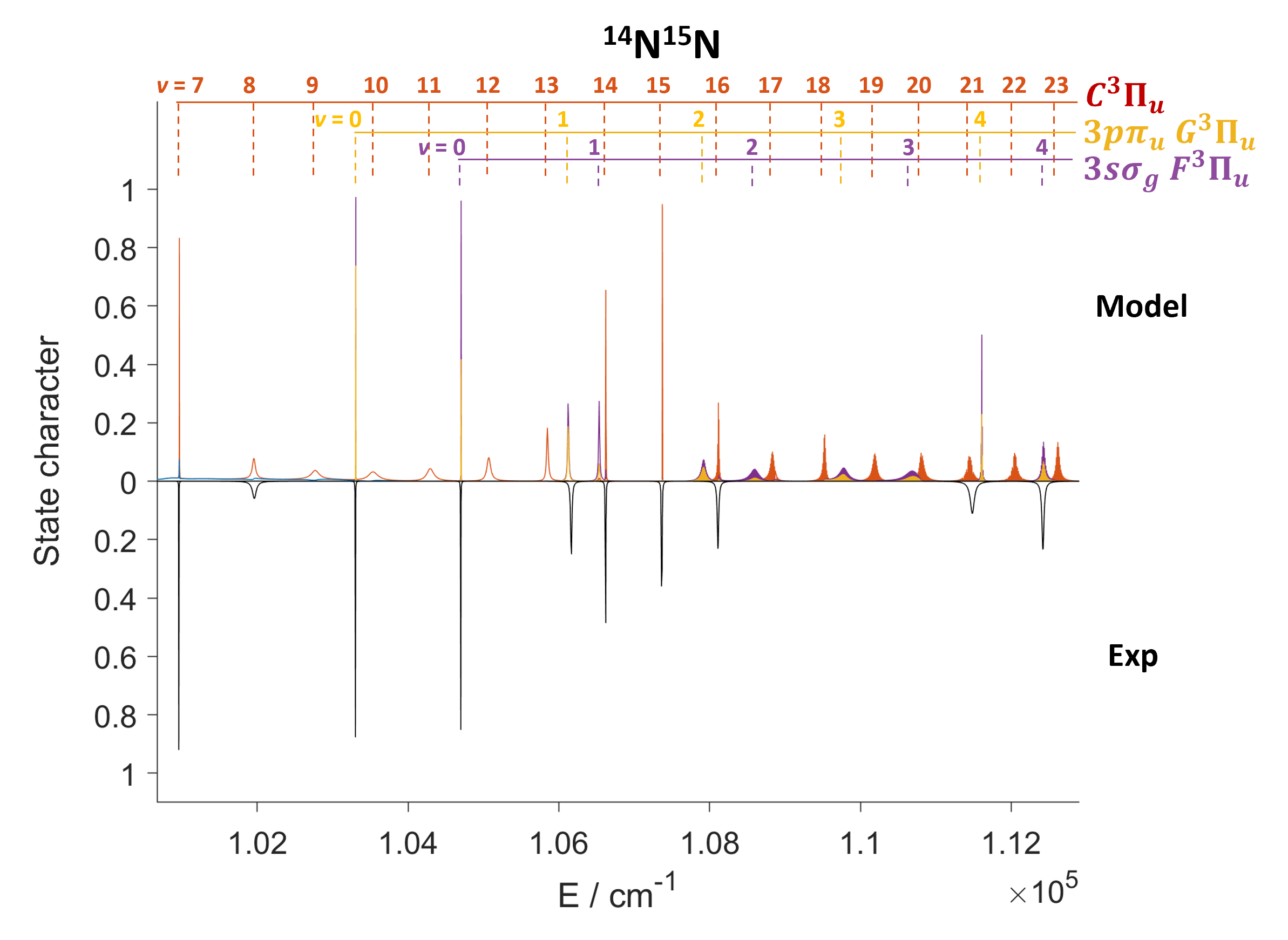}
\caption{Comparison of the valence-hole model (calculated at $J=1$) and the experimental observations for the $^3\Pi_u$ electronic states of $^{14}$N$^{15}$N. The experimental bound-state character plot in the lower panel is generated based on the experimental observations summarized in Table I of Ref~\citenum{heays2019spin} (level energy, $T_{00}$, and predissociation linewidth, $\Gamma$).}
\label{fig:n2_3piu_fit_1415}
\end{figure}

\begin{figure}
\includegraphics[width=3.5 in]{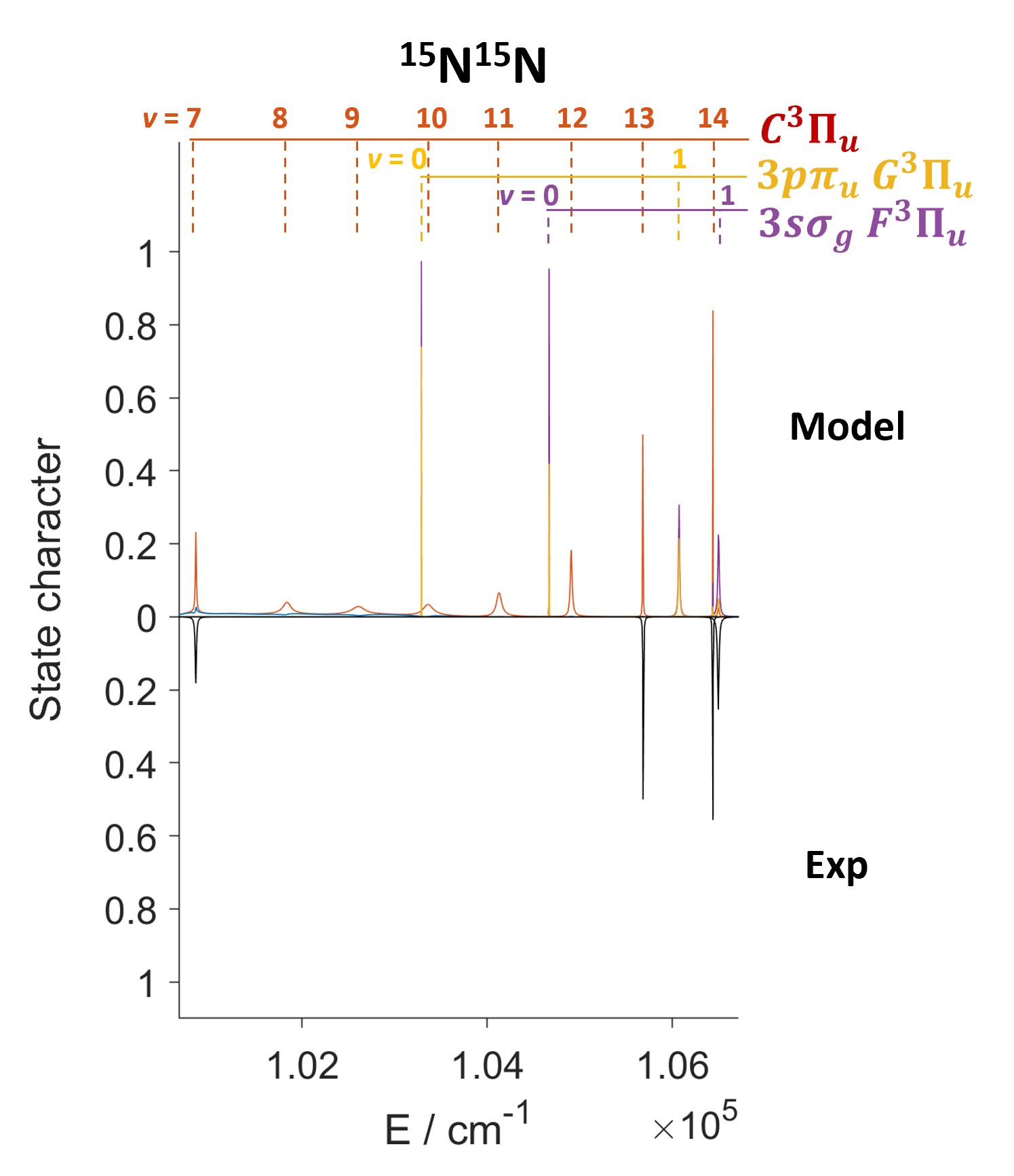}
\caption{Comparison of the valence-hole model (calculated at $J=1$) and the experimental observations for the $^3\Pi_u$ electronic states of $^{15}$N$_2$. The experimental bound-state character plot in the lower panel is generated based on the experimental observations summarized in Table I of Ref~\citenum{lewis2008coupled} (level energy, $T_{exp}$, and predissociation linewidth, $\Gamma_{exp}$).}
\label{fig:n2_3piu_fit_1515}
\end{figure}

The results of the experimental fit model for the $^3\Pi_u$ electronic states of $^{14}$N$^{15}$N and $^{15}$N$_2$ are shown, respectively, in Figs.~\ref{fig:n2_3piu_fit_1415} and \ref{fig:n2_3piu_fit_1515}, where the total bound diabatic state character is shown as a function of energy above the first dissociation limit, as in Fig.~3a of the main text.

\newpage

\newpage

\newpage




\section{Global diabatization schemes}
\label{sec:global_scheme}

\subsection{Valence-hole models of the $ab$ $initio$ calculations}
\label{sec:ab_initio_vh}

The molecular parameters used to diabatize the CASSCF potentials for the N$_2$ $^3\Pi_u$, Si$_2$ $^3\Pi_g$, and SiC $^3\Pi$ states are listed, respectively, in Tables~\ref{tab:valence_hole_N2}-\ref{tab:valence_hole_SiC}. The CASSCF potential energies are provided in a separate Excel document. The molecular parameters used to diabatize the $R$-matrix potentials of the N$_2$ $^3\Pi_u$ states from Ref~\citenum{little2013ab} are listed in Table~\ref{tab:valence_hole_N2_Rmatrix}. 

In Tables~\ref{tab:valence_hole_N2}-\ref{tab:valence_hole_N2_Rmatrix}, the lowest-energy dissociation channel for a specified electronic state system is taken to be the reference for the energies of various diabatic states (e.g., the electronic state term energy, $T_e$) and their associated separated-atom limits in the fit model. The ``$T_e$\,(Diss)” column gives the values of $T_e$ relative to that reference. The energies of each separated-atom limit from the $ab$ $initio$ calculation, relative to the same reference, are indicated in the ``Diss. Limit” column.

\subsection{Valence-hole models of the experimental observations}
\label{sec:exp_vh}

The molecular parameters used to fit the experimental observations for the C$_2$ $^3\Pi_g$, N$_2$ $^3\Pi_u$, and CN $^2\Sigma^+$ electronic states are listed, respectively, in Tables~\ref{tab:valence_hole_C2_exp}-\ref{tab:valence_hole_CN_exp}. As discussed in the Method section of the main text, in each case, an effective Hamiltonian in the diabatic basis is constructed and fitted to the experimental observations. 

For the C$_2$ $^3\Pi_g$ electronic state system, the vibronic term energies for the observed levels of its four electronic states are reproduced with a mean absolute error of 0.34\,cm$^{-1}$. The $B$ constants are reproduced with a mean absolute error of 0.0077\,cm$^{-1}$. For the fit, the ro-vibrational wavefunctions from each diabatic electronic state in the model are obtained using the DVR method~\cite{colbert1992novel}, with an evenly spaced grid of $R$ ($\Delta \textup{R}=0.01\,\textup{\AA}$) for $0.8\,\,\textup{\AA}\leq R\leq R_{\textup{max}}=7\,\textup{\AA}$. The values of the vibronic term energy and rotational $B$ constant for levels in the relevant energy region converge well with respect to $R_{\textup{max}}$, $\Delta \textup{R}$, and the number of basis states (100 for the d$_1$ diabatic state and 600 for the other four diabats, as defined in Table~\ref{tab:valence_hole_C2_exp}). To derive an approximate predissociation linewidth for the $e^3\Pi_g$ $v=12$ resonance from the fit model with a sufficiently high energy resolution, a much larger $R_{\textup{max}}$ of 198\,$\,\textup{\AA}$ is used for the simulation in Figs.~1d and 1e of the main text. The number of basis states used for that simulation has been increased accordingly.

For the CN $^2\Sigma^+$ electronic state system, the vibronic term energies for the observed levels of the $B^2\Sigma^+$ and $E^2\Sigma^+$ states are reproduced with a mean absolute error of 0.33\,cm$^{-1}$. The $B$ constants are reproduced with a mean absolute error of 0.0013\,cm$^{-1}$. The DVR ro-vibrational wavefunctions from each diabatic electronic state in the model are calculated with the same $R$-range, $R_{\textup{max}}$, and $\Delta \textup{R}$ values, as applied for the C$_2$ $^3\Pi_g$ state fit. For the Hamiltonian of the fit model for the CN $^2\Sigma^+$ states, 100 basis states are used for the d$_1$ state and 600 basis states for the other two diabats, as defined in Table~\ref{tab:valence_hole_CN_exp}. 

For the experimental fit model of the N$_2$ $^3\Pi_u$ states, the DVR wavefunctions of the ro-vibrational basis states from each diabatic electronic state (six valence + two Rydberg) are calculated with a $\Delta \textup{R}$ value of 0.01\,$\textup{\AA}$ for $0.8\,\,\textup{\AA}\leq R\leq R_{\textup{max}}=21\,\textup{\AA}$. The use of a relatively large $R_{\textup{max}}$ here, while computationally more expensive (by $\sim$$10\times$ compared to the use of $R_{\textup{max}}=7\,\textup{\AA}$), allows me to incorporate the observed predissociation linewidths to the fit. To increase computational efficiency during the non-linear fit, only the six lowest valence states from the fit model of the CASSCF (Table~\ref{tab:valence_hole_N2}) and $R$-matrix (Table~\ref{tab:valence_hole_N2_Rmatrix}) calculations are included in the experimental fit model. The exclusion of the three high-energy valence states is not expected to significantly impact the validity of the experimental fit model for the N$_2$ $^3\Pi_u$ states. For the Hamiltonian of this fit model, 250 basis states are used for the d$_1$ state and the two Rydberg states, and 1800 basis states for each of the other five diabats, as defined in Table~\ref{tab:valence_hole_N2_exp}. The molecular parameters for the d$_4$ and d$_6$ diabats are fixed to the values of the $R$-matrix fit model (Table~\ref{tab:valence_hole_N2_Rmatrix}). To increase energy resolution, a larger $R_{\textup{max}}$ of 105\,$\,\textup{\AA}$ is used for the bound-state character simulation in Fig.~3 of the main text, and in Figs.~\ref{fig:n2_3piu_fit_1415} and \ref{fig:n2_3piu_fit_1515} of the SI. 

The coupled-channel model of the N$_2$ $^3\Pi_u$ states by Lewis and co-workers~\cite{lewis2008coupled}, which differs from the valence-hole model here in the choice of the diabatic basis (as discussed in Section~3 of main text), provides a good reference during the initial stage of the experimental fit. In this coupled-channel model, the valence$\sim$valence ($C\sim C'$), valence$\sim$Rydberg ($C'\sim G$ and $C'\sim F$), and Rydberg$\sim$Rydberg ($G\sim F$) interactions are all assumed to be independent of $R$. The $H^{el}_{ij}$ electrostatic interaction terms for the d$_1$-d$_2$, d$_2$-d$_7$, and d$_2$-d$_7$ pairs in the valence-hole model (as defined in Table~\ref{tab:valence_hole_N2_exp}) are, respectively, the direct cause of the $C\sim C'$, $C'\sim G$, and $C'\sim F$ interactions in Lewis' model. The interaction between the d$_7$ and d$_8$ Rydberg states in Table~\ref{tab:valence_hole_N2_exp} is equivalent to the $G\sim F$ interaction in Lewis' work. The $C\sim C'$, $C'\sim G$, $C'\sim F$, and $G\sim F$ interaction matrix elements derived by Lewis and co-workers~\cite{lewis2008coupled} are good starting points for the corresponding terms in the fit. For simplicity, the d$_1$-d$_2$, d$_2$-d$_7$, d$_2$-d$_7$, and d$_7$-d$_8$ interactions are all assumed to be $R$-independent in the valence-hole fit model, as specified in Table~\ref{tab:valence_hole_N2_exp}.

\newpage

\begin{table} 
\caption{Molecular parameters used in the global diabatization scheme for the N$_2$ $^3\Pi_u$ states calculated by the CASSCF method (Fig.~2b of the main text). The exponential decay rates ($s_{ij}$) for the electrostatic interactions ($\mathcal{H}_{ij}e^{-s_{ij} R}$) are assumed to be the same for all ($s_{ij}=s=1.4130\,\text{\AA}^{-1}$) but one pair of diabatic states ($s_{26}=0.8347\,\text{\AA}^{-1}$), as defined in this table. In the last column, the 2s$^2$2p$^3$ label is omitted for all the LS states of atomic nitrogen with that electron configuration. The energy values here are all referenced to the ($^4S$)+($^2D$) separated-atom limit.}
\begin{center}
\begin{tabular}{@{\hspace{8pt}} c @{\hspace{8pt}} | @{\hspace{6pt}} c  @{\hspace{6pt}} | @{\hspace{6pt}} c   @{\hspace{6pt}}| c  @{\hspace{6pt}} c   @{\hspace{6pt}} | @{\hspace{6pt}} c   @{\hspace{6pt}}| c   @{\hspace{6pt}} c @{\hspace{6pt}} @{\hspace{6pt}} c   @{\hspace{6pt}}c  | @{\hspace{6pt}} c @{\hspace{6pt}} | @{\hspace{6pt}} c @{\hspace{6pt}}}

\hline
diabats & $T_e$\,(Diss)\,/\,eV & $\beta$\,/\,$\text{\AA}^{-1}$ & & $R_e\,/\,\text{\AA}$& $h$ & Diss. Limit\\
\hline
d$_1$& 0.4684 & 3.1809 & &	1.1550 & 5.83 & ($^4S$)+2s2p$^4$($^4P$)\\
 &  &  & &  &  & $Ref$\,+\,8.4542\,eV\\
d$_2$& 0.9285 & 5.5360 & & 1.9215 & 238.79 & ($^4S$)+($^2P$) \\
 &  &  & &  &  & $Ref$\,+\,0.9299\,eV\\
d$_3$& 1.6519 & 3.8242 & & 1.4915 & 46.35 & ($^2D$)+($^2D$) \\
 &  &  & &  &  & $Ref$\,+\,2.8764\,eV\\
d$_4$& 3.1846 & 3.8382 & & 1.5379 & 34.67 & ($^2D$)+($^2P$) \\
 &  &  & &  &  & $Ref$\,+\,3.8063\,eV\\
d$_5$& 3.7336 & 2.1852 & & 2.5884 & 2.86 & ($^2D$)+($^2P$) \\
 &  &  & &  &  & $Ref$\,+\,3.8063\,eV\\
\hline
 & $A_r$\,/\,eV & $k_r$\,/\,$\text{\AA}^{-1}$ & & &  & \\
\hline
d$_6$ & 555.061 & 2.8612 & &	&  & ($^4S$)+($^2D$) ($Ref$)\\
d$_7$ & 212.446 & 2.8339 & &	&  & ($^2D$)+($^2D$)\\
 &  &  & &  &  & $Ref$\,+\,2.8764\,eV\\
d$_8$ & 966.609 & 3.4226 & &	&  & ($^2D$)+($^2P$)\\
 &  &  & &  &  & $Ref$\,+\,3.8063\,eV\\
d$_9$ & 9174.83 & 4.5968 & &	&  & ($^2P$)+($^2P$)\\
 &  &  & &  &  & $Ref$\,+\,4.7362\,eV\\
\hline\hline
 interaction & $\mathcal{H}_{ij}$\,/\,eV & interaction && $\mathcal{H}_{ij}$\,/\,eV & interaction & $\mathcal{H}_{ij}$\,/\,eV\\
\hline
d$_1$-d$_2$ & 0.5636 & d$_1$-d$_5$&& 1.5493 &d$_1$-d$_8$& 6.0341 \\
d$_1$-d$_3$ & 7.2520 & d$_1$-d$_6$&& 3.4200 &d$_1$-d$_9$& 1.2846\\
d$_1$-d$_4$ & 2.0894 & d$_1$-d$_7$&& 5.0792 &d$_4$-d$_7$& 6.6714\\
\hline
d$_2$-d$_6$ & 3.5424 & && &&\\


\end{tabular}
\end{center}
\label{tab:valence_hole_N2}
\end{table}

\newpage

\begin{table} 
\caption{Molecular parameters used in the global diabatization scheme for the Si$_2$ $^3\Pi_g$ states calculated by the CASSCF method (Fig.~6c of the main text). The exponential decay rates ($s_{ij}$) for all five pairs of electrostatic interactions ($\mathcal{H}_{ij}e^{-s_{ij} R}$) are assumed to be the same. These decay rates are determined to be $s_{ij}=s=0.8707\,\text{\AA}^{-1}$ from the fit. In the last column, the 3s$^2$3p$^2$ label is omitted for all the LS states of atomic silicon with that electron configuration. The energy values here are all referenced to the ($^3P$)+($^3P$) separated-atom limit.}
\begin{center}
\begin{tabular}{@{\hspace{8pt}} c @{\hspace{8pt}} | @{\hspace{6pt}} c  @{\hspace{6pt}} | @{\hspace{6pt}} c   @{\hspace{6pt}}| c  @{\hspace{6pt}} c   @{\hspace{6pt}} | @{\hspace{6pt}} c   @{\hspace{6pt}}| c   @{\hspace{6pt}} c @{\hspace{6pt}} @{\hspace{6pt}} c   @{\hspace{6pt}}c  | @{\hspace{6pt}} c @{\hspace{6pt}} | @{\hspace{6pt}} c @{\hspace{6pt}}}

\hline
diabats & $T_e$\,(Diss)\,/\,eV & $\beta$\,/\,$\text{\AA}^{-1}$ & & $R_e\,/\,\text{\AA}$& $h$ & Diss. Limit\\
\hline
d$_1$& 1.0278 & 1.7495 & &	2.1989 & -0.77 & ($^3P$)+3s3p$^3$($^5S$)\\
 &  &  & &  &  & $Ref$\,+\,2.9384\,eV\\
d$_2$& -0.6843 & 2.5402 & & 2.4409 & 64.09 & ($^3P$)+($^3P$) ($Ref$)\\
d$_3$& -0.2464 & 1.5682 & & 2.5188 & 3.51 & ($^3P$)+($^1D$) \\
 &  &  & &  &  & $Ref$\,+\,1.0787\,eV\\
d$_4$& 0.5688 & 2.3748 & & 2.5587 & 57.23 & ($^3P$)+($^1D$) \\
 &  &  & &  &  & $Ref$\,+\,1.0787\,eV\\
d$_5$& 1.9013 & 1.7709 & & 3.8639 & 54.61 & ($^3P$)+($^1S$) \\
 &  &  & &  &  & $Ref$\,+\,1.9342\,eV\\
\hline
 & $A_r$\,/\,eV & $k_r$\,/\,$\text{\AA}^{-1}$ & & &  & \\
\hline
d$_6$ & 139.303 & 1.9970 & &	&  & ($^3P$)+($^1D$)\\
 &  &  & &  &  & $Ref$\,+\,1.0787\,eV\\
\hline\hline
 interaction & $\mathcal{H}_{ij}$\,/\,eV & interaction && $\mathcal{H}_{ij}$\,/\,eV & interaction & $\mathcal{H}_{ij}$\,/\,eV\\
\hline
d$_1$-d$_2$ & 3.3541 & d$_1$-d$_4$&& 0.9512 &d$_1$-d$_6$&3.0308\\
d$_1$-d$_3$ & 2.9057 & d$_1$-d$_5$&& 1.8144 &&\\

\end{tabular}
\end{center}
\label{tab:valence_hole_Si2}
\end{table}

\newpage

\begin{table} 
\caption{Molecular parameters used in the global diabatization scheme for the SiC $^3\Pi$ states calculated by the CASSCF method (Fig.~6d of the main text). The exponential decay rates ($s_{ij}$) are assumed to be the same for each pair of electrostatic interactions ($\mathcal{H}_{ij}e^{-s_{ij} R}$). These decay rates are determined to be $s_{ij}=s=0.9760\,\text{\AA}^{-1}$ from the fit. In the last column (C+Si), the 3s$^2$3p$^2$ label is omitted for all the LS states of atomic silicon with that electron configuration, and the 2s$^2$2p$^2$ label is omitted for all the LS states of atomic carbon with that electron configuration. The energy values here are all referenced to the C($^3P$)+Si($^3P$) separated-atom limit.}
\begin{center}
\begin{tabular}{@{\hspace{8pt}} c @{\hspace{8pt}} | @{\hspace{6pt}} c  @{\hspace{6pt}} | @{\hspace{6pt}} c   @{\hspace{6pt}}| c  @{\hspace{6pt}} c   @{\hspace{6pt}} | @{\hspace{6pt}} c   @{\hspace{6pt}}| c   @{\hspace{6pt}} c @{\hspace{6pt}} @{\hspace{6pt}} c   @{\hspace{6pt}}c  | @{\hspace{6pt}} c @{\hspace{6pt}} | @{\hspace{6pt}} c @{\hspace{6pt}}}

\hline
diabats & $T_e$\,(Diss)\,/\,eV & $\beta$\,/\,$\text{\AA}^{-1}$ & & $R_e\,/\,\text{\AA}$& $h$ & Diss. Limit\\
\hline
d$_1$ & 0.1161 & 3.0854 & &	1.7286 & 45.76 & ($^3P$)+3s3p$^3$($^5S$)\\
 &  &  & &  &  & $Ref$\,+\,2.9384\,eV\\
d$_2$ & -0.8576 & 3.6349 & & 1.9596 & 286.23 & ($^3P$)+($^3P$) ($Ref$) \\
d$_3$ & -0.1473 & 2.0304 & & 2.0621 & 6.24 & ($^3P$)+($^1D$) \\
 &  &  & &  &  & $Ref$\,+\,1.0787\,eV\\
d$_4$ & 0.7988 & 1.6428 & & 2.0909 & -2.86 & ($^1D$)+($^3P$) \\
 &  &  & &  &  & $Ref$\,+\,1.6242\,eV\\
d$_5$ & 2.7896 & 1.6$\,^a$ & & 3.0$\,^a$ & 10$\,^a$ & 2s2p$^3$($^5S$)+($^3P$) \\
 &  &  & &  &  & $Ref$\,+\,3.0872\,eV\\
\hline
 & $A_r$\,/\,eV & $k_r$\,/\,$\text{\AA}^{-1}$ & & &  & \\
\hline
d$_6$ & 114.608 & 2.1295 & &	&  & ($^1D$)+($^3P$)\\
 &  &  & &  &  & $Ref$\,+\,1.6242\,eV\\
d$_7$ & 99.609 & 1.8186 & &	&  & ($^1D$)+($^3P$)\\
 &  &  & &  &  & $Ref$\,+\,1.6242\,eV\\
d$_8$ & 496.863 & 2.3919 & &	&  & ($^3P$)+($^1S$)\\
 &  &  & &  &  & $Ref$\,+\,1.9590\,eV\\
d$_9$ & 5690.88 & 3.2* & &	&  & ($^1S$)+($^3P$)\\
 &  &  & &  &  & $Ref$\,+\,2.7277\,eV\\
d$_{10}$ & $\,^b$& $\,^b$& &	&  & ($^3P$)+($^1D$)\\
 &  &  & &  &  & $Ref$\,+\,1.0787\,eV\\
d$_{11}$ & $\,^c$& $\,^c$& &	&  & ($^3P$)+($^1D$)\\
 &  &  & &  &  & $Ref$\,+\,1.0787\,eV\\
\hline\hline
 interaction & $\mathcal{H}_{ij}$\,/\,eV & interaction && $\mathcal{H}_{ij}$\,/\,eV & interaction & $\mathcal{H}_{ij}$\,/\,eV\\
\hline
d$_1$-d$_2$ & 3.6576 & d$_1$-d$_5$&& 0.062$\,^a$ &d$_1$-d$_{10}$&0.9383\\
d$_1$-d$_3$ & 2.5099 & d$_1$-d$_6$&& 3.8448 &d$_1$-d$_{11}$&0.3003\\
d$_1$-d$_4$ & 1.8359 & d$_1$-d$_8$&& 0.3443 &&\\
\hline
d$_4$-d$_{10}$ & 1.5013 & d$_4$-d$_{11}$&& 0.4738 &d$_{10}$-d$_{11}$&0.6815\\


\end{tabular}
\end{center}
\footnotesize{$^a$ Fixed in the fit model.}\\
\footnotesize{$^b$ A bi-exponential decay function, $\mathcal{A}_{r}e^{-k_{r1} R}+\mathcal{A}_{r2}e^{-k_{r2} R}$, is used for d$_{10}$, where $\mathcal{A}_{r1}=180504$\,eV, $k_{r1}=7.2821$\,$\text{\AA}^{-1}$, $\mathcal{A}_{r2}=5.2154$\,eV, and $k_{r2}=0.9133$\,$\text{\AA}^{-1}$.}\\
\footnotesize{$^c$ The following potential is used for d$_{11}$, $E_{\infty}+\left[a\left(1-\frac{b}{e^{cR+d}}\right)^2+f\right]e^{-R}$, where $E_{\infty}$ is the energy of the ($^3P$)+($^1D$) dissociation limit; $a=0.6251$\,eV, $b=5508.13$, $c=2.7223$\,$\text{\AA}^{-1}$, $d=696.00$, and $f=0.9991$\,eV.}\\

\label{tab:valence_hole_SiC}
\end{table}

\begin{table} 
\caption{Molecular parameters used in the global diabatization scheme for the N$_2$ $^3\Pi_u$ states calculated by the $R$-matrix method (Fig.~\ref{fig:R_matrix_fit} of the SI). The exponential decay rates ($s_{ij}$) for the electrostatic interactions ($\mathcal{H}_{ij}e^{-s_{ij} R}$) are assumed to be the same for all ($s_{ij}=s=1.5385\,\text{\AA}^{-1}$) but one pair of diabatic states ($s_{26}=0.6816\,\text{\AA}^{-1}$), as defined in this table. In the last column, the 2s$^2$2p$^3$ label is omitted for all the LS states of atomic nitrogen with that electron configuration. The energy values here are all referenced to the ($^4S$)+($^2D$) separated-atom limit.}
\begin{center}
\begin{tabular}{@{\hspace{8pt}} c @{\hspace{8pt}} | @{\hspace{6pt}} c  @{\hspace{6pt}} | @{\hspace{6pt}} c   @{\hspace{6pt}}| c  @{\hspace{6pt}} c   @{\hspace{6pt}} | @{\hspace{6pt}} c   @{\hspace{6pt}}| c   @{\hspace{6pt}} c @{\hspace{6pt}} @{\hspace{6pt}} c   @{\hspace{6pt}}c  | @{\hspace{6pt}} c @{\hspace{6pt}} | @{\hspace{6pt}} c @{\hspace{6pt}}}

\hline
diabats & $T_e$\,(Diss)\,/\,eV & $\beta$\,/\,$\text{\AA}^{-1}$ & & $R_e\,/\,\text{\AA}$& $h$ & Diss. Limit\\
\hline
d$_1$ & -0.6666 & 2.7930 & &	1.1595 & 2.72 & ($^4S$)+2s2p$^4$($^4P$)\\
 &  &  & &  &  & $Ref$\,+\,8.5439\,eV\\
d$_2$ & -0.0802 & 2.8122 & & 1.5186 & 3.53 & ($^4S$)+($^2P$) \\
 &  &  & &  &  & $Ref$\,+\,1.1915\,eV\\
d$_3$ & 0.6276 & 3.0558 & & 1.5248 & 17.68 & ($^2D$)+($^2D$) \\
 &  &  & &  &  & $Ref$\,+\,2.3841\,eV\\
d$_4$ & 2.4388 & 3.8171 & & 1.5660 & 100$\,^{a}$ & ($^2D$)+($^2P$) \\
 &  &  & &  &  & $Ref$\,+\,3.5756\,eV\\
d$_5$ & 3.2645 & 2.9633 & & 2.1798 & 80$\,^{a}$ & ($^2D$)+($^2P$) \\
 &  &  & &  &  & $Ref$\,+\,3.5756\,eV\\
\hline
 & $A_r$\,/\,eV & $k_r$\,/\,$\text{\AA}^{-1}$ & & &  & \\
\hline
d$_6$ & 560.710 & 3.0864 & &	&  & ($^4S$)+($^2D$) ($Ref$)\\
d$_7$ & 89.494 & 2.6361 & &	&  & ($^2D$)+($^2D$)\\
 &  &  & &  &  & $Ref$\,+\,2.3841\,eV\\
d$_8$ & 145.175 & 2.7249 & &	&  & ($^2D$)+($^2P$)\\
 &  &  & &  &  & $Ref$\,+\,3.5756\,eV\\
d$_9$ & 500000$\,^{a}$  & 6.9213 & &	&  & ($^2P$)+($^2P$)\\
 &  &  & &  &  & $Ref$\,+\,4.7671\,eV\\
\hline
Rydberg & $T_e$\,(Diss)\,/\,eV & $\beta$\,/\,$\text{\AA}^{-1}$ & & $R_e\,/\,\text{\AA}$& $h$ & Diss. Limit\\
\hline
d$_{10}$($3p\,\pi_u$) & 0.8456 & 2.8693 & & 1.1202 & 1.31 & ($^4S$)+2s$^2$2p$^2$3p($^4D$)\\
 &  &  & &  &  & $Ref$\,+\,9.3721\,eV\\
d$_{11}$($3s\,\sigma_g$) & 1.0702 & 3.1010 & & 1.1852 & 7.63 & ($^4S$)+2s$^2$2p$^2$3s($^4P$)\\
 &  &  & &  &  & $Ref$\,+\,7.9463\,eV\\
\hline\hline
 interaction & $\mathcal{H}_{ij}$\,/\,eV & interaction && $\mathcal{H}_{ij}$\,/\,eV & interaction & $\mathcal{H}_{ij}$\,/\,eV\\
\hline
d$_1$-d$_2$ & 2.1251 & d$_1$-d$_5$&& 1.6391 &d$_1$-d$_8$& 5.7072 \\
d$_1$-d$_3$ & 8.3235 & d$_1$-d$_6$&& 2.7538 &d$_1$-d$_9$& 5.5925\\
d$_1$-d$_4$ & 3.7765 & d$_1$-d$_7$&& 4.9384 &d$_4$-d$_7$& 9.6606 \\
\hline
d$_2$-d$_6$ & 3.2659 & && &&\\
\hline
d$_{10}$-d$_{11}$ & 1.0869 & d$_3$-d$_{10}$&& 0.3789 &d$_4$-d$_{11}$& 1.0416 \\
d$_2$-d$_{10}$ & 1.0721 & d$_3$-d$_{11}$&& 0.8914 &d$_7$-d$_{10}$& 1.5071\\
d$_2$-d$_{11}$ & 0.9353 & d$_4$-d$_{10}$&& 1.6979 &d$_7$-d$_{11}$& 1.2128 \\


\end{tabular}
\end{center}
\footnotesize{$^a$ Fixed in the fit model.}\\
\label{tab:valence_hole_N2_Rmatrix}
\end{table}

\begin{table} 
\caption{Molecular parameters of the experimental fit model for the C$_2$ $^3\Pi_g$ states (Fig.~1 of the main text). Numbers in parentheses are $1\sigma$ uncertainties of the last digits. In the last column, the 2s$^2$2p$^2$ label is omitted for all the LS states of atomic carbon with that electron configuration. The ($^3P$)+($^3P$) separated-atom limit, which is taken as the reference for all of the relevant dissociation limits here, is 50391 cm$^{-1}$ above the $v=0$, $J=0$ level of the $X^1\Sigma_g^+$ state.}
\begin{center}
\begin{tabular}{@{\hspace{8pt}} c @{\hspace{8pt}} | @{\hspace{6pt}} c  @{\hspace{6pt}} | @{\hspace{6pt}} c   @{\hspace{6pt}}| c  @{\hspace{6pt}} c   @{\hspace{6pt}} | @{\hspace{6pt}} c   @{\hspace{6pt}}| c   @{\hspace{6pt}} c @{\hspace{6pt}} @{\hspace{6pt}} c   @{\hspace{6pt}}c  | @{\hspace{6pt}} c @{\hspace{6pt}} | @{\hspace{6pt}} c @{\hspace{6pt}}}

\hline
valence & $T_e$\,/\,cm$^{-1}$ & $\beta$\,/\,$\text{\AA}^{-1}$ & & $R_e\,/\,\text{\AA}$& $h$ & Diss. Limit\\
\hline
d$_1$ & 23699.6(756) & 2.7928(464) & & 1.2688(27) & 8.19(124) & ($^3P$)+2s2p$^3$($^5S$)\\
 &  &  & &  &  & $Ref$\,+\,33735.2\,cm$^{-1}$\\
d$_2$ & 41016.9(362) & 2.9617(558) & & 1.6128(28) & 13.07(436) & ($^3P$)+($^3P$) ($Ref$)\\
d$_3$ & 40882.8(1317) & 2.3063(338) & & 1.5682(45) & 6.46(131) & ($^3P$)+($^1D$) \\
 &  &  & &  &  & $Ref$\,+\,10192.7\,cm$^{-1}$\\
d$_4$ & 51293.7(735) & 2.9873(186) & & 1.6302(48) & 30$\,^a$ & ($^3P$)+($^1D$) \\
 &  &  & &  &  & $Ref$\,+\,10192.7\,cm$^{-1}$\\
\hline
 & $A_r$\,/\,cm$^{-1}$ & $k_r$\,/\,$\text{\AA}^{-1}$ & & &  & \\
\hline
d$_5$ & 2846994.8$\,^{b}$ & 3.1273$\,^{b}$ & &	&  & ($^3P$)+($^1D$)\\
 &  &  & &  &  & $Ref$\,+\,10192.7\,cm$^{-1}$\\
\hline
\hline
$s_{ij}$\,/\text{\AA}$^{-1}$ & interaction & $\mathcal{H}_{ij}$\,/\,cm$^{-1}$ && interaction & $\mathcal{H}_{ij}$\,/\,cm$^{-1}$ & \\
\hline
1$\,^a$ & d$_1$-d$_3$ & 35290.1(2782) && d$_1$-d$_4$  & 12370.6(5440) &\\
        & d$_1$-d$_5$ & 18710.2$\,^{b}$ && d$_2$-d$_3$  & 14856.7(2766) &\\


\end{tabular}
\end{center}
\footnotesize{$^a$ Fixed in the fit model.}\\
\footnotesize{$^b$ Fixed to the values from the fit model of the MRCI potentials, assuming $s_{ij}=1\text{\AA}^{-1}$ for all pairs of $H_{ij}^{el}$}\\
\label{tab:valence_hole_C2_exp}
\end{table}

\begin{table} 
\caption{Molecular parameters of the experimental fit model for the N$_2$ $^3\Pi_u$ states (Figs.~2c and 3 of the main text). Numbers in parentheses are $1\sigma$ uncertainties of the last digits. In the last column, the 2s$^2$2p$^3$ label is omitted for all the LS states of atomic nitrogen with that electron configuration. The ($^4S$)+($^2D$) separated-atom limit, which is taken as the reference for all of the relevant dissociation limits here, is 97938 cm$^{-1}$ above the $v=0$, $J=0$ level of the $X^1\Sigma_g^+$ state.}
\begin{center}
\begin{tabular}{@{\hspace{8pt}} c @{\hspace{8pt}} | @{\hspace{6pt}} c  @{\hspace{6pt}} | @{\hspace{6pt}} c   @{\hspace{6pt}}| c  @{\hspace{6pt}} c   @{\hspace{6pt}} | @{\hspace{6pt}} c   @{\hspace{6pt}}| c   @{\hspace{6pt}} c @{\hspace{6pt}} @{\hspace{6pt}} c   @{\hspace{6pt}}c  | @{\hspace{6pt}} c @{\hspace{6pt}} | @{\hspace{6pt}} c @{\hspace{6pt}}}

\hline
valence & $T_e$\,/\,cm$^{-1}$ & $\beta$\,/\,$\text{\AA}^{-1}$ & & $R_e\,/\,\text{\AA}$& $h$ & Diss. Limit\\
\hline
d$_1$ B& 93438.4(127) & 2.8225(10) & &	1.1508 & 2.78 & ($^4S$)+2s2p$^4$($^4P$)\\
 &  &  & &  &  & $Ref$\,+\,68911\,cm$^{-1}$\\
d$_2$ H& 99978.3(17) & 3.0343(17) & & 1.5020 & 7.42 & ($^4S$)+($^2P$) \\
 &  &  & &  &  & $Ref$\,+\,9610\,cm$^{-1}$\\
d$_3$ A& 104793.8(43) & 3.3699(12) & & 1.5268 & 57.89 & ($^2D$)+($^2D$) \\
 &  &  & &  &  & $Ref$\,+\,19229\,cm$^{-1}$\\
d$_4$ C& 118783.6$\,^a$ & 3.8171$\,^a$ & & 1.5660$\,^a$ & 100$\,^a$ & ($^2D$)+($^2P$) \\
 &  &  & &  &  & $Ref$\,+\,28839\,cm$^{-1}$\\
\hline
 & $A_r$\,/\,cm$^{-1}$ & $k_r$\,/\,$\text{\AA}^{-1}$ & & &  & \\
\hline
d$_5$ I& 5017788 & 3.0864$\,^a$ & &	&  & ($^4S$)+($^2D$) ($Ref$)\\
d$_6$ D& 721817$\,^a$ & 2.6361$\,^a$ & &	&  & ($^2D$)+($^2D$)\\
 &  &  & &  &  & $Ref$\,+\,19229\,cm$^{-1}$\\
\hline
Rydberg & $T_e$\,/\,cm$^{-1}$ & $\beta$\,/\,$\text{\AA}^{-1}$ & & $R_e\,/\,\text{\AA}$& $h$ & Diss. Limit\\
\hline
d$_{7}$($3p\,\pi_u$) & 104052.4(63) & 2.4996(49) & & 1.1164 & -0.59 & ($^4S$)+2s$^2$2p$^2$3p($^4D$)\\
 &  &  & &  &  & $Ref$\,+\,75591\,cm$^{-1}$\\
d$_{8}$($3s\,\sigma_g$) & 105159.2(20) & 2.8409(27) & & 1.1789 & 3.37 & ($^4S$)+2s$^2$2p$^2$3s($^4P$)\\
 &  &  & &  &  & $Ref$\,+\,64091\,cm$^{-1}$\\
\hline
\hline
$s_{ij}$\,/\text{\AA}$^{-1}$ & interaction & $\mathcal{H}_{ij}$\,/\,cm$^{-1}$ && interaction & $\mathcal{H}_{ij}$\,/\,cm$^{-1}$ & Note \\
\hline
0$\,^b$ & d$_1$-d$_2$ & 1108.6(51) && d$_7$-d$_8$ & 1160.5(36) & $R$-independent \\
 & d$_2$-d$_7$ & 1065.8(202) && d$_2$-d$_8$  & 366.3(80)  &  $H_{ij}^{el}$\\
\hline
1.5385$\,^a$ & d$_1$-d$_3$ & 61144.0(863) && d$_1$-d$_4$ & 30459.5$\,^a$ & \\
 & d$_1$-d$_6$ & 39830.8$\,^a$ && d$_4$-d$_6$ &  77918.2$\,^a$ & \\
 & d$_3$-d$_7$ & 2921.2 && d$_3$-d$_8$ & 4819.7 & \\
\hline
1$\,^b$ & d$_2$-d$_5$ & 38382.9 &&  &&\\


\end{tabular}
\end{center}
\footnotesize{$^a$ Fixed to the values from the fit model of the $R$-matrix potentials. $^b$ Fixed in the fit model.}\\
\label{tab:valence_hole_N2_exp}
\end{table}

\begin{table} 
\caption{Molecular parameters of the experimental fit model for the CN $^2\Sigma^+$ states (Fig.~4 of the main text). Numbers in parentheses are $1\sigma$ uncertainties of the last digits. In the last column (C+N), the 2s$^2$2p$^2$ label is omitted for all the LS states of atomic carbon with that electron configuration, and the 2s$^2$2p$^3$ label is omitted for all the LS states of atomic nitrogen with that corresponding electron configuration. The C($^3P$)+N($^2D$) separated-atom limit, which is taken as the reference for all of the relevant dissociation limits here, is 81640.2 cm$^{-1}$ above the $v=0$, $J=0$ level of the $X^2\Sigma^+$ state.}
\begin{center}
\begin{tabular}{@{\hspace{8pt}} c @{\hspace{8pt}} | @{\hspace{6pt}} c  @{\hspace{6pt}} | @{\hspace{6pt}} c   @{\hspace{6pt}}| c  @{\hspace{6pt}} c   @{\hspace{6pt}} | @{\hspace{6pt}} c   @{\hspace{6pt}}| c   @{\hspace{6pt}} c @{\hspace{6pt}} @{\hspace{6pt}} c   @{\hspace{6pt}}c  | @{\hspace{6pt}} c @{\hspace{6pt}} | @{\hspace{6pt}} c @{\hspace{6pt}}}

\hline
valence & $T_e$\,/\,cm$^{-1}$ & $\beta$\,/\,$\text{\AA}^{-1}$ & & $R_e\,/\,\text{\AA}$& $h$ & Diss. Limit\\
\hline
d$_1$ & 34288.5(10541) & 2.0489(643) & & 1.1453(32) & -0.37(22) & 2s2p$^3$($^3D$)+($^2D$)\\
 &  &  & &  &  & $Ref$\,+\,64089.9\,cm$^{-1}$\\
d$_2$ & 52022.2(582) & 2.1081(128) & & 1.3815(14) & -0.73(10) & ($^3P$)+($^2D$) ($Ref$)\\
d$_3$ & 80012.8(9241) & 2.5$\,^a$ & & 1.6$\,^a$ & 0$\,^a$  & ($^3P$)+($^2D$) ($Ref$)\\
\hline
\hline
$s_{ij}$\,/\text{\AA}$^{-1}$ & interaction & $\mathcal{H}_{ij}$\,/\,cm$^{-1}$ && interaction & $\mathcal{H}_{ij}$\,/\,cm$^{-1}$ & \\
\hline
1$\,^a$ & d$_1$-d$_2$ & 44130.0(6049) && d$_1$-d$_3$  & 49137(5073) &

\end{tabular}
\end{center}
\footnotesize{$^a$ Fixed in the fit model.}\\

\label{tab:valence_hole_CN_exp}
\end{table}


\clearpage

\bibliography{vh_JPCA_supplemental}